# Microelectromechanical Resonators for Radio Frequency Communication Applications

*Review Paper*[*]

Joydeep Basu[1] and Tarun Kanti Bhattacharyya[2]

Department of Electronics and Electrical Communication Engineering

Indian Institute of Technology Kharagpur, Kharagpur 721302, West Bengal, India

Email: [1]joydeepkgp [at] gmail.com, [2]tkb [at] ece.iitkgp.ernet.in

**Abstract:** Over the past few years, microelectromechanical system (MEMS) based on-chip resonators have shown significant potential for sensing and high frequency signal processing applications. This is due to their excellent features like small size, large frequency-quality factor product, low power consumption, low cost batch fabrication, and integrability with CMOS IC technology. Radio frequency communication circuits like reference oscillators, filters, and mixers based on such MEMS resonators can be utilized for meeting the increasing count of RF components likely to be demanded by the next-generation multi-band/multi-mode wireless devices. MEMS resonators can provide a feasible alternative to the present-day well-established quartz crystal technology that is riddled with major drawbacks like relatively large size, high cost, and low compatibility with IC chips. This article presents a survey of the developments in this field of resonant MEMS structures with detailed enumeration on the various micromechanical resonator types, modes of vibration, equivalent mechanical and electrical models, materials and technologies used for fabrication, and the application of the resonators for implementing oscillators and filters. These are followed by a discussion on the challenges for RF MEMS technology in comparison to quartz crystal technology; like high precision, stability, reliability, need for hermetic packaging etc. which remain to be addressed for enabling the inclusion of micromechanical resonators into tomorrow's highly integrated communication systems.

---

# 1. Introduction

Wireless communication technologies have witnessed major advances since the late 1980s. In the present wireless communication scenario, numerous standards such as CDMA (code-division multiple access), GSM (global system for mobile communication), the emerging 3G (3rd generation) and 4G exist, which provide us with voice, data and broadband communication. But, in order to maintain the quality and reliability of these state of the art technologies, the specifications given to a communication circuit design engineer are getting more and more rigorous. The continual adoption of such stringent requirements as demanded by advanced wireless systems such as software defined radios and cognitive radio systems requires development of emerging technologies such as radio frequency microelectromechanical system (RF MEMS) based devices (De Los Santos et al. 2004). RF MEMS switches, varactors, inductors, and resonators are ideal for reconfigurable systems at GHz range of operation. These components normally possess low insertion loss and very high quality factor (Q) even up to tens of GHz frequency. Moreover, RF MEMS devices generate very low intermodulation products. Thus, with the availability of high performance passives over wide frequency range of operation, and the immense potential of integration with high volume industry mainstream complementary metal-oxide-semiconductor (CMOS) electronics; RF MEMS based circuits have become an important technology of interest to the wireless design fraternity (Tilmans et al. 2003; Brand and Fedder 2005; Kim and Chun 2007).

A ubiquitous component in modern integrated electronic systems is a frequency-reference circuit. Piezoelectric quartz crystals offering a large quality factor (> 100,000), superior frequency-temperature stability (< 35 ppm over 0 to 70ºC), reliability, technical maturity, and extensive commercial availability are presently in high demand for use in such frequency selection and clocking applications. Since 2000, the requirement of quartz crystals and crystal oscillators has been growing steadily between 4 to 10 % annually fuelled by the tremendous growth in demand for various electronic equipments (Lam 2008). But, resonators based on quartz do have some limitations. These cannot be miniaturized easily for on-chip usage, involves a costly manufacturing process, and their performance degrades when subjected to severe levels of shock and vibration. Again, the on-chip tank circuits with monolithic inductors and capacitors provided by the present-day integrated circuit technology suffer from poor Q-values (~ 10) (Lin et al. 2004b). Thus, MEMS based resonators have emerged as an attractive alternative which can offer Q-values close to that for quartz in both vacuum as well as in air and operating frequencies up to the very-high frequency (VHF, 30 MHz–300 MHz) and ultra-high frequency (UHF, 300

MHz–3 GHz) ranges, consume less power, provide temperature stability better than 18 ppm over 27 to 107ºC and aging stability better than 2 ppm over one year, have shorter design and production cycle times, and can be monolithically integrated and fabricated using low-cost CMOS compatible processes (Nguyen 2007). In addition, MEMS resonators are very robust to shock and vibration, and provide an overwhelming size advantage. However, for replacement of the quartz crystal technology with MEMS, the latter has to successfully overcome a few problems related to temperature stability, thermal hysteresis, long-term stability, packaging etc. These issues are gradually being solved, and the timing and frequency control industry have started embracing the MEMS resonator based devices (Yole 2010).

Both capacitive (Hao et al. 2004; Lin et al. 2004b; Wang et al. 2004a) and piezoelectrically transduced (Stephanou and Pisano 2006; Zuo et al. 2010) versions of micromechanical resonators have been demonstrated. Various applications encompassing such resonators, including RF filtering and mixing have also been reported (Wong and Nguyen 2004). Some prior surveys on MEMS resonators are available in the literature. An excellent review by Nguyen (2007) focused on the various aspects of RF MEMS resonators, filters, and reference oscillators. Silicon micromachined resonators have also been reviewed in the paper on RF MEMS technology by Kim and Chun (2007). Lam (2008) provided a comprehensive review of MEMS and quartz crystal technologies used in frequency control applications from an industrial perspective. This paper expands on the subject by reviewing the latest developments, while simultaneously providing a systematic account of the different geometries, vibration modes, mechanical and electrical equivalent models, fabricating materials, and RF applications of such resonators; and hence can serve as a tutorial for this field of micro-scale electromechanical resonators.

Capacitively transduced MEMS resonators are in general favored than their piezoelectrically transduced counterparts. Also among the former variety of microresonators, bulk acoustic mode of vibration is the preferred option for realizing high frequency of operation (Clark et al. 2005; Hao et al. 2004). So, besides a description of the various reported MEMS resonator types and modes of vibration, this review provides a dedicated focus on the operation and modeling of a disk-resonator which belongs to the particular variety of bulk-mode resonator based on capacitive transduction. But at the same time, this suffices for understanding of the principle behind flexural-mode resonators as well. The materials and methods used by various research groups for fabrication of MEMS resonators have also been elaborated with the corresponding advantages and disadvantages. The loss-mechanisms responsible for dissipation of the energy of vibration of such resonators have also been discussed. Light has also been shed on some of the reported applications of RF MEMS resonant devices, like oscillators and filters. Furthermore, the prospects of integration of MEMS resonators with CMOS circuitry have been briefly reviewed. Finally, some future research directions have been provided which should enable the incorporation of such resonators into the mainstream of electronic communication devices.

## 2. Modes of Vibration of MEMS Resonators

Idea about a micromechanical resonator can be easily made from the simple macro-scale example of a guitar string of a particular length and made of nickel and steel alloy which vibrates at a distinct resonance frequency when plucked. Hence, it mechanically selects a particular frequency with a Q of about 350 which is far better than passive electronic resonators. Now if the dimensions of such a string are scaled down to the micron level, and fabricated with IC-compatible materials like silicon, polysilicon etc., and excited electrostatically or piezoelectrically rather than mere plucking, a microscopic-guitar is created, named *clamped-clamped* (CC) beam resonator (illustrated in Table 1) in MEMS terminology, which can provide us with a resonance frequency of about 10 MHz at a Q of about 10,000 (Nguyen 2007). This is one of the simplest versions of MEMS resonators. An even simpler and perhaps the foremost example is a surface-micromachined cantilever beam fixed at one end used as a resonator, called the *resonant gate transistor* (RGT) as shown in Fig. 1 (Nathanson et al. 1967). Hence, such a MEMS resonator essentially consists of the following three components: (1) an input-transducer which converts the input electrical signal into a mechanical signal i.e., an electrostatic force (alternatively piezoelectric, magnetostatic etc.), (2) a mechanical resonant structure which can vibrate in one or more modes due to the produced electrostatic force, and (3) an output-transducer that senses the motion of the vibrating structure, hence converting the mechanical signal back to an output electrical signal.

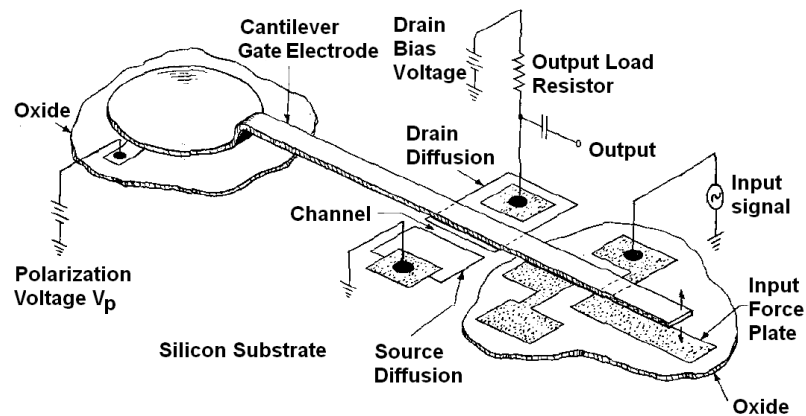

**Fig. 1.** Schematic diagram of a resonant gate transistor consisting of a gold beam which is 0.1 mm in length, and 5–10 μm in thickness, resonating at 5 kHz with a quality factor of 500. [Reprinted with permission from (Nathanson et al. 1967). Copyright (1967) IEEE.]

Furthermore, a *free-free* (FF) beam resonator can be realized by supporting the beam at the nodes rather than at the ends for minimizing anchor losses, hence leading to improved performance.

Micromechanical resonators can have different shapes like beams, square plates, circular disks, annular rings, comb etc., and can again be classified according to their modes of operation, namely *flexural*, *torsional,* and *bulk* mode devices (Fig. 2(a)) (Taylor and Huang 1997; Chandorkar et al. 2008):

- Flexural mode of vibration is representative of transverse standing waves. In such devices, the displacement of the structures is orthogonal to the bending stress in the structure.
- In resonators vibrating in the torsional mode, the dominant stress is shear-stress and the displacement produced is rotational in nature.
- Bulk mode operation can be described in terms of standing longitudinal waves.

Fig. 2(b) shows the commonly used vibrational modal shapes for a square-plate bulk mode device: *extensional* (or, *contour*), *wine-glass*, and *Lamé*. The modal shapes of the first flexural mode of CC and FF beams are provided in Fig. 2(c).

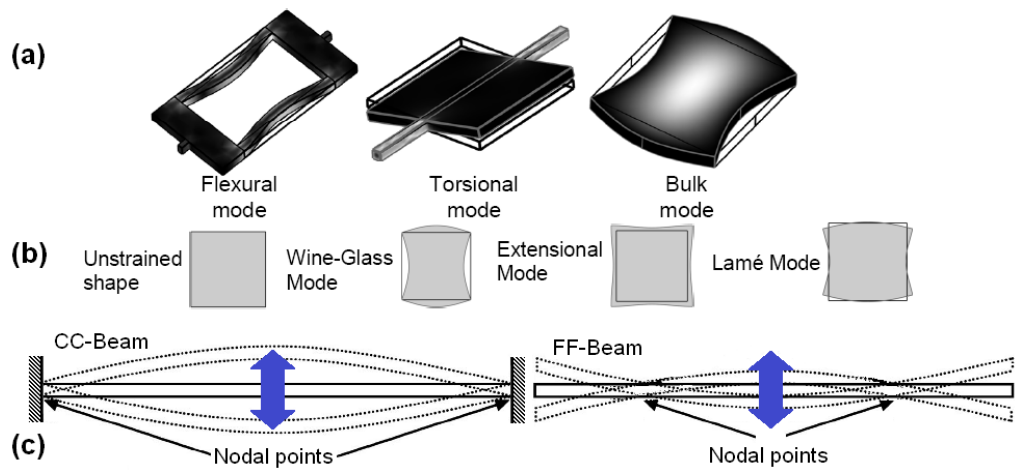

**Fig. 2.** (a) Three basic types of vibration of mechanical resonators: flexural, torsional and bulk mode structures. (b) Various mode shapes for a bulk mode square plate resonator [Reprinted with permission from (Chandorkar et al. 2008). Copyright (2008) IEEE]. (c) First flexural-mode of resonance for CC and FF beams.

In general, bulk-mode vibration of microresonators is preferred for high frequency generation due to larger structural stiffness in comparison to other modes. Moreover, bulk-modes yield higher Q relative to flexural mode resonators of the same frequency. This is due to the fact that flexural modes have comparatively larger surface-to-volume ratios than bulk mode resonators, thus leading to increased losses from surface effects (Hao et al. 2004; Lee and Seshia 2009).

Hence, bulk-mode operation will be emphasized in this paper. A common example of a bulk-mode device is a circular-disk resonator which can vibrate in two distinct modes as illustrated in Fig. 3: (a) *radial-contour* (or, *extensional* or, *breathing*) mode where the shape of the disk expands and contracts equally in all the lateral surface, and (b) *elliptical* (or, *wine-glass*) mode where the disk expands along one axis and contracts in the orthogonal axis forming two alternate and perpendicular ellipses per cycle of vibration with four nodal points at the perimeter. Radial-contour modes provide higher values of effective stiffness (Wang et al. 2004a) and hence, are favored and will be discussed in detail in section 5.

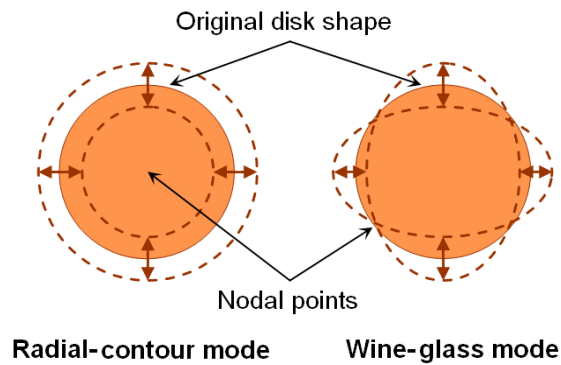

**Fig. 3.** The two modes of bulk acoustic vibration of a disk (in the first or, fundamental mode). The dotted curves show the fully expanded and contracted mode shapes.

The most common transduction type used in MEMS resonators is electrostatic actuation and sensing, as it can produce small components that are robust, relatively simple to fabricate with materials compatible with integrated transistor circuits, have better thermal stability, and are tolerant to environmental changes (Lucyszyn 2004). Capacitively transduced devices, in general offer the best frequency-Q products, since the signal transduction occurs without the need for direct physical contact between the electrodes and the resonating body, and thus suffer less from material-interface losses and quality factor degradation that can impede other transducer types (Nguyen 2007). Thus, devices based on this particular transduction mechanism will be emphasized in the following section.

## 3. Capacitively Transduced MEMS Resonator Structures

Various shapes of capacitive micromechanical resonators have been reported till date, operating in different vibration modes, and fabricated with different materials. Table 1 provides an expansive list of such devices.

One of the simplest MEMS resonator structure is a clamped-clamped (CC) beam fixed to the substrate at both ends, the idea of which probably originated from the reporting of miniature cantilevers and doubly supported mechanical beams by Howe and Muller (1983), fabricated from polycrystalline silicon using the conventional MOS planar process. Bannon et al. (2000) have demonstrated a 2 μm thick polysilicon flexural mode CC-beam resonator with Q ~ 8,000 at a frequency of 10 MHz in vacuum ambient, and an equivalent motional resistance ($R_e$, its significance will be discussed later on) of about 5 kΩ. The Q gets reduced to ~ 50 at 10 MHz when measured in air, and also gets reduced due to anchor losses as frequency increases e.g., Q ~ 300 at 70 MHz in vacuum. A similar device by Lin et al. (2004b) is illustrated in row 1 of Table. 1. Single-crystal silicon (SCS) has also been used in making CC-beam resonators, as by Pourkamali et al. (2003) (second entry in Table 1). They have used the *HARPSS* (High Aspect Ratio combined Poly and Single crystal Silicon) process, where the beam is of SCS, while the drive and sense electrodes are made of polysilicon. Here, vertical capacitive gaps ($d_0$) as small as 80 nm and measured Q as high as 177,000 for a 19 kHz beam have been demonstrated under 1 mtorr vacuum. Recently, Nabki et al. (2011) have reported 2 μm thick CC-beams fabricated using amorphous silicon carbide (SiC) with transducer gaps of 100 nm; exhibiting resonant frequencies up to 26.2 MHz and Q in the order of 1000 when tested at pressures below 200 μtorr. However, scaling the CC-beam to higher frequencies (~ 50 MHz) quickly degrades the performance due to energy dissipation to the substrate via their anchors limiting the Q to values below 500. This can be somewhat minimized in an FF-beam resonator (Table 1) by supporting it at the flexural mode nodal locations using four quarter-wavelength support beams, with Q remaining high even with frequencies increasing past 100 MHz (Wang et al. 2000). However, as the dimensions of beam resonators are further scaled down to achieve higher resonant frequencies; the attainable Q tends to decrease due to increased surface-loss and anchor-loss mechanisms (Hao et al. 2004).

One of the earliest resonator designs is the laterally driven electrostatic *comb-drive* structure reported in 1989 by Tang et al., fabricated with polysilicon and having a Q of about 100 at resonant frequency of 40 kHz (Nguyen 1995; Tang et al. 1989). A comb-drive is a common actuator used in MEMS and consists of rows of interlocking teeth; half of the teeth are attached to a fixed beam and the other half attach to a movable beam assembly. By applying the same polarity voltage to both parts, the resultant electrostatic force repels the movable beam away from the fixed, and vice-versa. In this manner the comb-drive can be moved "in" or "out" and either dc or ac voltages can be applied. Here, there is no pull-in effect between the electrodes and the displacement is linear, contrary to straight actuation. Cioffi and Hsu (2005) have reported a silicon comb-drive resonator using *SOI MEMS* process (Table 1), with Q as high as 50,000.

Square plates operating in the bulk acoustic mode of vibration is another common resonator structure (Table 1). Lee et al. (2008) have reported a Q of $1.16 \times 10^6$ in their silicon extensional mode device vibrating at 2.18 MHz, excited through lateral capacitive gap drive electrodes on each side of the structure. Poly SiC has been utilized by Bhave et al. (2005) in making such square plate resonators. Another widely reported structure is a circular disk, anchored at one or more nodal points on its periphery (for wine-glass mode), or supported by a stem at its center (for radial-contour mode), as shown in corresponding schematics in Table 1. Lin et al. have achieved sub 100 nm capacitive gaps in their elliptic bulk-mode disk resonators of polysilicon, exhibiting a motional resistance of only 1.5 kΩ (Abdelmoneum et al. 2003; Lin et al. 2004a). Pourkamali et al. (2004) have reported HARPSS-based 18 μm thick SCS disk resonators with a Q of 46,000 at an elliptical mode resonance frequency of 150 MHz, exhibiting a resistance of 43.3 kΩ for 160 nm capacitive gaps. Recently, a silicon resonator of similar type but with a larger radius by Lee and Seshia (2009) has shown an exceptional Q of $1.9 \times 10^6$. A radial-contour mode disk of polysilicon has been reported by Clark et al. (2005), with a Q of about 10,000 at 156 MHz. Reports on micromechanical resonators based on flexural vibrations of circular disks (Huang 2008) and square plates (Demirci and Nguyen 2006) can also be found in literature (Table 1). However, flexural mode resonators reach the scaling limits around 100 MHz, as the dimensions required for higher frequencies become difficult to achieve. Also, the motional impedance grows rapidly. Then, bulk acoustic mode resonators come to the rescue by getting around these limitations with the employment of resonant modes with much higher effective stiffness, increasing the operating frequency.

Ring resonators consisting of a circular ring supported by spokes emanating from a stem anchor at its center have also been reported, with the ring expanding and contracting in width (Table 1). Quarter-wavelength dimensioning of spokes nulls losses to the stem anchor, allowing this design to achieve large Q's beyond 1 GHz (Li et al. 2004). A similar square-shaped device, but operating in flexural mode has been commercially unveiled by SiTime Corp. of Sunnyvale, Calif. in 2006 as the SiT0100 part (Highbeam Research 2006; Kim and Chun 2007; Tabatabaei and Partridge 2010). An evaluation of a SiTime MEMS resonator with respect to a comparable quartz resonator is given in Table 2. In 2003, Discera Inc. of San Jose, Calif. was the first to offer a 19.2 MHz all silicon MEMS oscillator for multiband wireless transceivers, based on a 30 μm × 8 μm beam resonator.

**Table 1.** Some reported capacitively transduced MEMS resonator structures*

| Type & Ref. | Material | Dimension(s) | Freq. and Q | Features | Schematic diagram |
|---|---|---|---|---|---|
| Clamped-clamped beam (flexural mode) (Lin et al. 2004b) | Poly silicon (2 μm thick) | Beam length = 40 μm, width = 8 μm | 9.34 MHz & 3,100 | Motional resistance, $R_e$ = 8.27 kΩ, $d_o$ = 100 nm, $V_p$ = 8 V | |
| Clamped-clamped beam (flexural mode) (Pourkamali et al. 2003) | Single crystal silicon (20 μm thick) | Beam length = 700 μm, width = 6 μm | 80 kHz & 74,000 | 80 nm to submicron transduction gaps ($d_o$) | |
| | | Beam length = 200 μm, width = 10 μm | 3.2 MHz & 4,500 | | |
| Free-free beam (flexural mode) (Wang et al. 2000) | Poly silicon (2.05 μm thick) | Beam length = 13.1 μm, width = 6 μm, supporting beam length = 10.3 μm, width = 1 μm | 92 MHz & 7,450 | $R_e$ = 167 kΩ, $V_p$ = 76 V | |
| Comb drive (flexural mode) (Cioffi and Hsu 2005) Fig. from Nguyen 1995 | Single crystal silicon (30 μm thick) | No. of comb fingers = 500, finger length = 10 μm, finger overlap = 4 μm | 32 kHz & 50,000 | $R_e$ = 420 kΩ, $V_p$ = 2.5 V, 1 μm gap between static and moving fingers | |
| Square plate extensional (bulk acoustic mode) (Lee et al. 2008) | Single crystal silicon (25 μm thick) | Side length = 2 mm | 2.18 MHz & 1,160,000 | $V_p$ = 60 V, $d_o$ = 3 μm | |
| Square plate lame (bulk acoustic) (Bhave et al. 2005) | Poly silicon carbide (2 μm thick) | Side length ≈ 35 μm | 173 MHz & 9,300 | $R_e$ = 18 kΩ, $d_o$ = 195 nm | |
| Wine-glass disk (bulk acoustic mode) (Lin et al. 2004a; Lee and Seshia 2009) Fig. from Abdelmoneum et al. 2003 | Poly silicon (3 μm thick) | Disk radius = 32 μm | 60 MHz & 48,000 | $R_e$ = 1.5 kΩ, $V_p$ = 12 V, $d_o$ = 80 nm | |
| | Single crystal silicon (25 μm thick) | Disk radius = 400 μm | 5.43 MHz & 1,900,000 | $R_e$ = 17 kΩ, $V_p$ = 60 V, $d_o$ = 2.7 μm | |
| Radial-contour disk (bulk acoustic) (Clark et al. 2005) | Poly silicon (2 μm thick) | Disk radius = 16.7 μm | 156 MHz & 9,290 | $R_e$ = 100 kΩ, $V_p$ = 35 V, $d_o$ = 100 nm | |

| Circular disk (flexural mode) (Huang 2008) | Nickel (3 μm thick) | Disk radius = 15 μm | 11.6 MHz & 1,651 | $R_e$ = 22 kΩ, $V_p$ = 5 V, $d_o$ = 200 nm | |
|---|---|---|---|---|---|
| Square plate (flexural mode) (Demirci and Nguyen 2006) | Poly silicon (2.2 μm thick) | Side length = 16 μm | 68 MHz & 15,000 | $R_e$ = 13.5 kΩ, $V_p$ = 25 V, $d_o$ = 90 nm | |
| Ring (contour mode) (Li et al. 2004) | Poly silicon (2 μm thick) | Radii: $r_i$ =11.8 μm, and $r_o$ =18.7 μm | 1.2 GHz & 15,000 | $R_e$ = 274 kΩ, $V_p$ = 10 V, $d_o$ = 100 nm | |
| Square (flexural mode) (Tabatabaei and Partridge 2010) | Single crystal silicon | Commercial model *SiT0100* from *SiTime*; size: 0.8 mm × 0.6 mm, 0.15 mm thick | 5.1 MHz & 80,000 | $V_p$ = 4.6 V, price ≈ $ 0.25 | |
| Triangular beam (torsional mode) (Naito et al. 2010) | Single crystal silicon | Beam length = 100 μm, width ($w_r$) = 4.25 μm, height ($t_r$) = 3 μm | 20 MHz & 220,000 | $R_e$ = 12 kΩ, $d_o$ = g = 130 nm, $V_{dc}$ = 1 V | |

\* Here, $R_e$ = motional resistance, $V_p$ = $V_{dc}$ = dc bias voltage, $d_0$ = transduction gap. Figures are reprinted with permission from respective references, and are copyrights of corresponding publishers.

**Table 2.** A comparison of quartz crystal and SiTime's silicon MEMS resonator (Kim and Chun 2007)

| Features | Quartz crystal resonator | MEMS resonator |
|---|---|---|
| Size | 2–5 mm | 400 μm |
| Frequency | 1–80 MHz | 1–50 MHz |
| Resonant Q (× $10^3$) | 100–200 | 75–150 |
| CMOS integration | No | Yes |
| Packaging | Ceramic or metal | Plastic |
| Aging (in the first year) | 3–5 ppm | 3 ppm |
| Compensated temperature stability | 1–10 ppm | 1–10 ppm |
| Shock/vibration immunity | Poor | Good |
| Cost | Higher | Lower |

Recently, an SiGe thin-film packaged SOI-based MEMS resonator has been revealed by Naito et al. (2010), exhibiting a huge quality factor of 220,000 at 20 MHz resonant frequency.

Their innovation consists of a triangular beam operated in a torsional vibration mode (Table 1) that enables lower anchor-loss and squeeze-film damping in comparison to flexural mode resonators. The polysilicon capacitive-gap electrodes have been aligned along the {111} side-planes of the SCS beam. The dc-bias requirement has been scaled down to a value of just 1 V, while exhibiting a reasonable motional resistance of 11.92 kΩ.

## 4. Materials Used in Fabrication of Capacitively Transduced MEMS Resonators

Researchers have reported the usage of different structural materials for fabricating the disk, stem, and I/O electrodes of disk resonators, a list of which is provided in Table 3. Thus, all such variations give changes in respective performances. A pertinent example is that of a disk resonator with material-mismatched stem (Wang et al. 2004b). Here, by employing different materials for fabricating the disk (polydiamond) and the supporting central stem (polysilicon), an acoustic impedance mismatch can be obtained at the disk-stem interface that suppresses the energy transfer from the vibrating disk to the stem. Hence, anchor-losses are reduced leading to substantially higher Q-values.

**Table 3.** Some reported radial-contour mode disk resonators made of different materials

| Material | Dimensions | Freq | Q | Features* | Ref. |
|---|---|---|---|---|---|
| Poly silicon | Disk radius = 10 μm, thickness = 2.1 μm, transducer gap = 68 nm, stem radius = 0.8 μm | 1.16 GHz | 2,683 | $V_p$ = 10.5 V, $R_e$ = 2442 kΩ, $L_e$ = 902 mH, $C_e$ = $2.1 \times 10^{-20}$ F | Wang et al. 2004a |
| Single crystal silicon | Disk radius = 100 μm, thickness = 20 μm, transducer gap = 110 nm | 24 MHz | 53,000 | $V_p$ = 5 V, $R_e$ = 2.1 kΩ, $L_e$ = 740 mH, $C_e$ = 60 aF, static capacitance $C_0$ = 0.6 pF, parallel parasitic cap. $C_{PP}$ = 1.2 pF | Sworowski et al. 2009 |
| Polydiamond disk with polySi stem | Disk radius = 10 μm, thickness = 3 μm, transducer gap = 90 nm, stem radius = 0.8 μm | 1.51 GHz | 11,555 | $V_p$ = 2.5 V, $R_e$ = 1.21 MΩ | Wang et al. 2004b |

* $V_p$ = dc bias voltage, $R_e$ = motional resistance, $L_e$ = motional inductance, and $C_e$ = motional capacitance.

Besides structural dimensions, the resonant frequency of a MEMS device is proportional to the acoustic velocity, which is $\sqrt{E/\rho}$, where E and ρ are Young's modulus and density respectively. In this regard, an assessment of the key properties of materials commonly used in fabricating microsystems seems pertinent which is presented in Table 4. Thus, for the same type of structure and dimensions, the polydiamond device will have the highest resonant frequency while the nickel devices will have the lowest. On the other hand, nickel and SiGe (silicon germanium) with their very low deposition temperatures seem to be promising from the post-transistor integration perspective of MEMS with CMOS technology utilizing low permittivity (or, low-k) dielectrics (Huang 2008; Quevy et al. 2006). But, as evident from the literature, silicon resonators (both poly and single crystalline) have attracted the maximum attention due to well-established processing technology available in IC industry and their compatibility with CMOS integration. Also, single-crystal silicon (SCS) has superior mechanical properties and hence, can offer resonators with lower internal losses as compared to those made from polysilicon. An SOI (silicon-on-insulator) based process can also provide with increased thickness of the structural layer, leading to increased transducer area; hence reduced motional resistance and higher power handling capability. However, as the conventional chemical vapor deposition (CVD) techniques cannot be used for depositing SCS, fabricating SCS resonators with ultrathin capacitive transduction gaps becomes difficult. Instances of MEMS resonators fabricated on alternative base materials like glass, polymer, ceramic etc. are relatively scarce than those on silicon substrate, as the later favors integration with CMOS technology (Jia and Madou 2005).

**Table 4.** Physical properties of popular MEMS structural materials (Huang 2008; Huang et al. 2006; Sze 2003)

| Material | Young's modulus $E$ (GPa) | Density $\rho$ (kg m$^{-3}$) | Poisson's ratio $\sigma$ | Deposition temperature (ºC) | Electrical conductivity ($10^7$ $\Omega^{-1}$ m$^{-1}$) |
|---|---|---|---|---|---|
| Silicon <100> | 130 | 2,330 | 0.28 | 1,000 | 0.00023 |
| Polysilicon | 150 | 2,300 | 0.226 | 588 | 0.001 |
| Polydiamond | 1,144 | 3,500 | 0.069 | 800 | 0.001 |
| Silicon carbide | 415 | 3,200 | 0.192 | 800 | 0.00083 |
| PolySi$_{0.35}$Ge$_{0.65}$ | 146 | 4,280 | 0.23 | 450 | 0.005 |
| Nickel | 195 | 8,900 | 0.31 | 50 | 1.43 |

## 5. Capacitively Transduced Circular Disk Resonator

A simple and commonly reported MEMS resonator geometry working on capacitive transduction is the circular-disk resonator which can provide us with a large quality factor at very high resonance frequencies (Basu and Bhattacharyya 2011). Thus, the theory and modeling of this particular variety of microresonators is discussed at length in this section, although the same principle is applicable for all other capacitively transduced resonator structures.

## 5.1 Disk Resonator Structure and Operation

A two-port disk resonator consists of a circular disk (made of say, polysilicon) suspended by a narrow cylindrical stem at its center. The centrally anchored disk resonator structure is actually favorable than a side-supported version where the disk is clamped to the anchors by using one or more thin support-beams attached to the lateral edge of the disk. This is due to the fact that the central point is actually a nodal-point for the radial-contour bulk-mode vibrations. Hence, usage of anchors at the disk periphery leads to some energy loss of the vibration through the anchors, called *anchor-loss* which will be discussed in section 6. Nonetheless, side-supported structures are the desirable geometry for disks vibrating in the wine-glass mode where nodal-points appear at the disk boundary. The disk is surrounded by lateral capacitive-gap input and output electrodes along the perimeter, which act as electromechanical transducers. The schematic diagram of such a disk resonator is given in Fig. 4(a). A cross-sectional view for the same is provided in Fig. 4(b) which illustrates the various material layers that can be utilized in fabricating the structure. Thus, the *Poly1* layer of polysilicon can be used as the structural layer for making the disk, stem and electrodes, while the *Poly0* can be patterned for taking electrical connections. Also, a deposition of *Nitride* (or, silicon dioxide) is necessary for acting as an insulating layer above the substrate. The *Metal* film is generally used for realizing pads and interconnects.

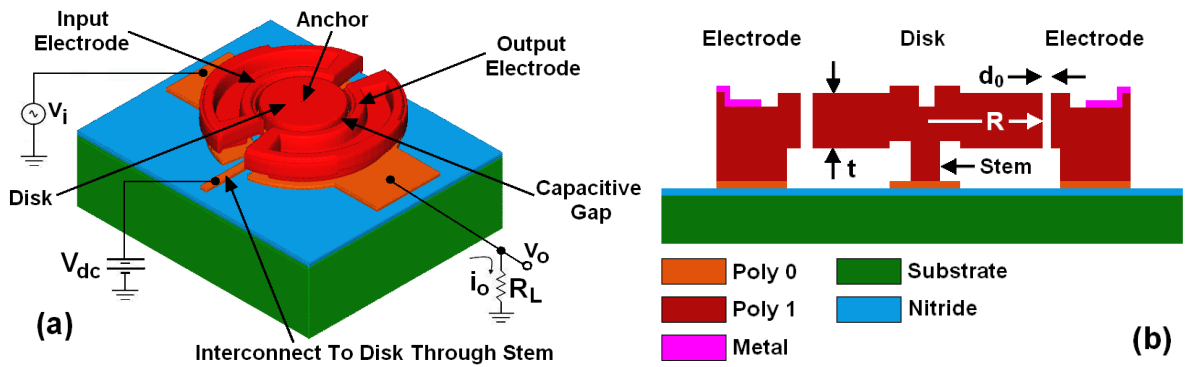

**Fig. 4.** (a) Schematic diagram of the disk resonator. (b) The cross-sectional view illustrating the various material layers used in the fabrication of this polysilicon resonator by means of surface micromachining.

As electrostatic force shows a quadratic dependence on the charge or the voltage, so the electrostatic transducers display a non-linear behavior. It can be made to behave linearly for incremental or small-signal variations around bias or equilibrium levels by applying a dc bias voltage, or by introducing some bias charge. The resonator under consideration can be excited into forced-vibrations by applying a sinusoidal voltage $v_i$ at the disk's mechanical resonance frequency $f_0$ to the input electrode and a dc-bias voltage $V_{dc}$ to the disk through its stem, hence producing a time-varying radial electrostatic force $F_i$ on the disk. Due to this force, there occurs a symmetrical expansion and contraction of the disk around its perimeter, and hence, a change in the disk-to-output electrode capacitance (which is dc-biased by $V_{dc}$) with time, thus producing a sinusoidal motional current $i_o$ with frequency $f_0$ in the load connected at the output port. Thus, an electrical signal at the input port gets transduced into a mechanical signal (force on the resonator) which is filtered by the high-Q mechanical response of the disk giving sinusoidal displacement of it at $f_0$. This mechanical response is translated back to the electrical domain by the output transducer (Clark et al. 2005; Wang et al. 2004a). Since the dc voltage $V_{dc}$ only charges the disk-to-electrode capacitor, no power consumption is incurred. The generated electrostatic force $F_e$ has both constant ($F_0$) and time varying components ($F_i$) in the radial direction, and is given by the relation:

$$F_e = \frac{1}{2}\left(\frac{\partial C_1}{\partial r}\right)(V_{dc} - v_i)^2 = F_0 + F_i \qquad (1)$$

Here $C_k$ denotes the electrode-to-resonator capacitance at the $k^{th}$ port, where k = 1 and 2 represent the input and output ports respectively. The time varying force $F_i$ is given in (2), where only the dominant term at resonance has been retained.

$$F_i \cong -V_{dc}\left(\frac{\partial C_1}{\partial r}\right)v_i \quad \text{(for, } V_{dc} >> v_i) \qquad (2)$$

$\partial C_1/\partial r$ is the change of electrode to resonator overlap capacitance per unit radial displacement at the input port, an expression for which can be obtained as in (3).

$$C_1(r) = C_0\left(1 - \frac{r}{d_0}\right)^{-1}$$

$$\Rightarrow \frac{\partial C_1}{\partial r} = \frac{C_0}{d_0}\left(1 - \frac{r}{d_0}\right)^{-2} \qquad (3)$$

where $C_0$ is the static drive-electrode-to-disk capacitance and is given by:

$$C_0 \cong \frac{\varepsilon_0 \phi R t}{d_0} \qquad (4)$$

Here, R and t are the radius and thickness of the disk respectively, $\varepsilon_0$ is the permittivity in vacuum, $d_0$ is the static electrode-to-resonator gap spacing, and $\phi$ is the angle defined by the edges of the input electrode (which is also same for the output electrode). Now, the following approximate expression can be derived:

$$\frac{\partial C_1}{\partial r} \cong \frac{\varepsilon_0 \phi R t}{d_0^2} \tag{5}$$

The zero to peak radial displacement amplitude at any point (r,θ) of the disk due to its radial-contour mode vibrations is given by (Wang et al. 2004a):

$$D(r, \theta) = AhJ_1(hr) \tag{6}$$

and at the perimeter (i.e., for r = R) by:

$$D(R, \theta) = \frac{QF_i}{jk_{re}} \tag{7}$$

Here, Q is the quality factor of the resonator, $k_{re}$ is the effective stiffness at the perimeter of the disk, A is a drive force-dependent ratio, and h is a constant as stated in (8).

$$h = \sqrt{(\omega_0^2 \rho) \bigg/ \left( \frac{E}{1+\sigma} + \frac{E\sigma}{1-\sigma^2} \right)} \tag{8}$$

The radial vibration of the disk creates a dc-biased time varying capacitance between the disk and the output electrode, which causes an output motional current, the expression for which can be written as follows:

$$i_0 = V_{dc} \frac{\partial C_2}{\partial t} \tag{9}$$

Eqn. (9) shows that the output current is generated only if a finite dc-bias voltage $V_{dc}$ is applied between the disk and the electrode. Therefore the device can be effectively turned off by switching off the bias voltage. This equation can be further modified using (2) and (6) yielding (10).

$$i_0 = -\frac{Q\omega_0}{k_{re}} \left( \frac{\partial C_1}{\partial r} \right) \left( \frac{\partial C_2}{\partial r} \right) V_{dc}^2 v_i \tag{10}$$

Thus, the output current is proportional to the input ac excitation voltage, and has the same frequency $\omega_0$. Although the two-port excitation configuration depicted in Fig. 4 generates an asymmetric force on the disk due to the ac-input being applied to only one side-electrode, symmetrical radial-contour modal-shape can be ensured as long as the input is at the breathing mode resonance frequency. The configuration of the I/O-electrodes also has a role in determining the phase relationship between the applied input voltage and the output current. Thus, for in-phase displacement of the resonator toward the sense and drive electrodes (e.g., for the two-electrode scheme of extensional bulk-mode disk resonator), the output current is 180º out of phase with respect to the input voltage; while for out-of-phase displacement of the resonator toward the sense and drive electrodes (e.g., for the four-electrode configuration of elliptical bulk-mode disk), the current coming out of the device is in-phase with the input voltage (Hao et al. 2004).

## 5.2 Disk Resonance Frequency

As cylindrical-coordinate system is used for describing the circular-disk resonator geometry, its resonance frequency for the radial-contour bulk vibrations can be derived in terms of Bessel functions (Clark et al. 2005; Hao and Ayazi 2007) as given by:

$$\frac{J_0\left(\frac{\varsigma}{\xi}\right)}{J_1\left(\frac{\varsigma}{\xi}\right)} = (1-\sigma) \tag{11}$$

$$\text{where, } \varsigma = \omega_0 R \sqrt{\frac{\rho(2+2\sigma)}{E}}, \quad \xi = \sqrt{\frac{2}{1-\sigma}} \tag{12}$$

Here $J_\alpha$ is Bessel function of the first kind of order α, $\omega_0$ (= $2\pi f_0$) is the angular resonance frequency, R is the radius of the disk, and E, ρ and σ are respectively the Young's modulus, mass density and Poisson's ratio of the material of the disk. For polysilicon, E ≈ 150 GPa, ρ ≈ 2,300 kg/m$^3$ and σ = 0.226. Above relations can be simplified to obtain the following expression for the resonant frequency for the i$^{th}$ breathing mode (Clark et al. 2005; Hao and Ayazi 2007):

$$f_0 = \frac{\omega_0}{2\pi} = \frac{\lambda_i}{2\pi R} \sqrt{\frac{E}{\rho(1-\sigma^2)}} \tag{13}$$

where $\lambda_i$ is the frequency parameter for that particular mode, the values of which are provided in Table 5 for the first four radial-contour modes.

**Table 5.** Values of $\lambda_i$ for the first four radial-contour modes

| Mode number (i) | Frequency parameter ($\lambda_i$) |
|---|---|
| 1 | 1.99 |
| 2 | 5.37 |
| 3 | 8.42 |
| 4 | 11.52 |

A plot of the resonance frequencies of the first four modes of radial-contour vibrations for varying disk-radius is shown in Fig. 5. Thus, lower disk-radius should be used for achieving higher resonant frequencies. The expressions of resonance frequency for some other common MEMS resonator structures are given in Table 6 for reference.

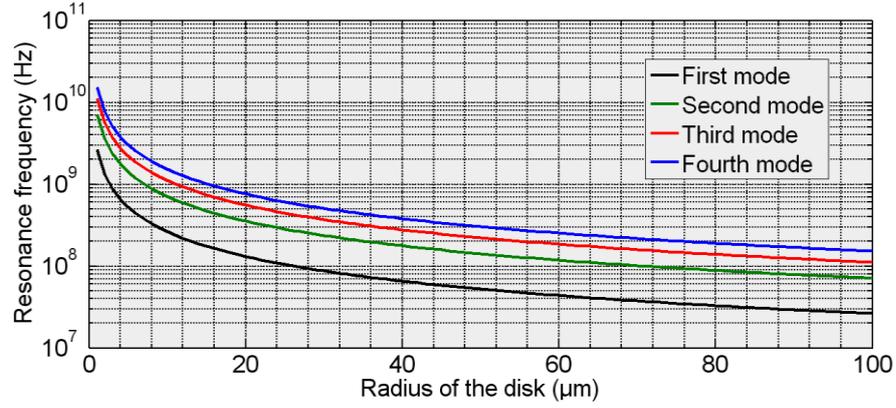

**Fig. 5.** Variation of radial-bulk mode resonance frequency with radius of the disk.

**Table 6.** Resonance frequency expressions of common micromechanical resonators*

| Type | Resonance Frequency ($f_0$) | Ref. |
|---|---|---|
| Clamped-clamped beam (flexural mode) | $$f_0 = 1.03\kappa\sqrt{\frac{E}{\rho}}\frac{h}{L_r^2}\left[1-\left(\frac{k_e}{k_m}\right)\right]^{1/2}$$ $$\frac{k_e}{k_m} = \int_{L_1}^{L_2}\frac{V_P^2\varepsilon_0 W_r}{[d(y)]^3 k_m(y)}dy$$ $$L_1 = 0.5(L_r - W_e),\ L_2 = 0.5(L_r + W_e)$$ *where $h$ = beam thickness, $L_r$ = beam length, $\kappa$ = scaling factor that models the effect of surface topography.* | Lin et al. 2004b |
| Square plate extensional (bulk mode) | $$f_0 = \sqrt{\frac{E}{4\rho L^2}}$$ *where $L$ = side length of the square plate.* | Lee et al. 2008 |
| Square plate (flexural mode) | $$f_0 = \frac{20.56}{2\pi}\frac{h}{L^2}\sqrt{\frac{E}{12\rho(1-\sigma^2)}}$$ *where $h$ = thickness of the plate, $L$ = length of one of its sides.* | Demirci and Nguyen 2006 |
| Wine-glass disk (bulk mode) | $$\left[\Psi_n\left(\frac{\zeta}{\xi}\right) - n - q\right][\Psi_n(\zeta) - n - q] = (nq - n)^2$$ $$\Psi_n(x) = \frac{xJ_{n-1}(x)}{J_n(x)},\ q = \frac{\zeta^2}{2n^2 - 2},$$ $$\zeta = 2\pi f_0 R\sqrt{\frac{2\rho(1+\sigma)}{E}},\ \xi = \sqrt{\frac{2}{1-\sigma}},\ n = 2$$ *where $J_n(x)$ = Bessel function of first kind of order n, $R$ = radius of the disk.* | Lin et al. 2004b |
| Circular ring (bulk contour mode) | $$f_0 = \frac{n}{4L_s}\sqrt{\frac{E}{\rho}}$$ $[J_1(hr_i)\sigma - J_1(hr_i) + r_i hJ_0(hr_i)] \times [Y_1(hr_o)\sigma - Y_1(hr_o) + r_o hY_0(hr_o)]$ $-[Y_1(hr_i)\sigma - Y_1(hr_i) + r_i hY_0(hr_i)] \times [J_1(hr_o)\sigma - J_1(hr_o) + r_o hJ_0(hr_o)] = 0$ *where $L_s$ = length of the radial support beams, $r_i$ = inner radius of the ring, $r_o$ = outer radius of the ring, $J_0\ (J_1)$ and $Y_0\ (Y_1)$ are Bessel functions of the first and second kinds respectively.* | Li et al. 2004 |

\* $E$ = Young's modulus, $\rho$ = density, and $\sigma$ = Poisson ratio of the structural material.

## 5.3 Equivalent Mechanical Model

Any MEMS resonator can be represented by an equivalent mechanical lumped-element model of a mass-spring-damper system having a single degree-of-freedom as depicted in Fig. 6. For arriving at the expressions for the effective mass ($m_{eff}$), spring-stiffness ($k_{eff}$) and damping factor ($b_{eff}$) of the model for a disk resonator, the expression of the total kinetic energy ($E_k$) of the vibrating disk should be utilized:

$$E_k = \frac{1}{2}\rho t \int_0^R \int_0^{2\pi} r u^2(r) \, dr \, d\theta \qquad (14)$$

Here, u(r,θ) denotes the velocity at any point (r,θ) on the disk. Thus, expression for the effective mass of the vibrating disk is as follows:

$$m_{eff} = \frac{2E_k}{u^2(R)} = \frac{2\pi\rho t \int_0^R r J_1(hr)^2 \, dr}{J_1(hR)^2} = \pi\rho t R^2 \left[1 - \frac{J_0(hR)J_2(hR)}{J_1(hR)^2}\right] \qquad (15)$$

$$\text{with, } h = \omega_0 \sqrt{\frac{\rho}{\left(\frac{E}{1+\sigma}\right) + \left(\frac{E\sigma}{1-\sigma^2}\right)}} = \frac{\lambda_i}{R} \qquad (16)$$

From Rayleigh's energy method (Schmidt and Howe 2008):

$$\omega_0 = \sqrt{k_{eff}/m_{eff}} \qquad (17)$$

This gives the relation between the effective spring-stiffness of the resonator and the resonance frequency $\omega_0$. Also, the damping factor that takes into account the energy-losses of the system is given by:

$$b_{eff} = \frac{\omega_o m_{eff}}{Q} = \frac{\sqrt{k_{eff} m_{eff}}}{Q} \qquad (18)$$

Micromechanical resonators have a low damping-loss which is articulated by their high Q-values.

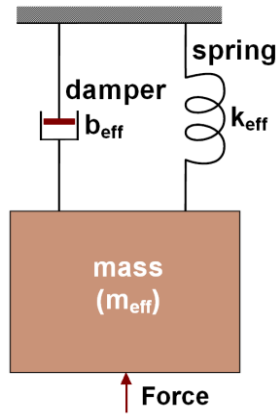

**Fig. 6.** An equivalent mechanical model for the resonator.

## 5.4 Equivalent Electrical Model

An electrical lumped-element model for a micromechanical resonator is significant from an electronic circuit simulation point of view. This can be obtained by replacing the mass, spring and dashpot elements of the mechanical model with their electrical analogs; which is possible due to the formal similarities of the integro-differential equations governing the behavior of mechanical and electrical systems. Using *force-voltage analogy* (Tilmans 1996) where force, displacement and velocity are analogous to voltage, charge and current respectively; the resonators can be assumed to exhibit series resonance. This is due to the fact that here, a through-variable (e.g., force) is equated to an across-variable (voltage), and vice-versa. So, a parallel arrangement in the mechanical domain is mapped to a series connection in the electrical equivalent circuit, and vice-versa. Now, once the expressions for the effective mass ($m_{eff}$), spring constant ($k_{eff}$) and viscous-damping coefficient ($b_{eff}$) are obtained as shown previously, a series-resonant RLC electrical model can be proposed for the microdisk resonator (Clark et al. 2005; Hao et al. 2004; Wang et al. 2004a) as depicted in Fig. 7(a), the elements of which are as follows:

$$l_e = m_{eff}$$
$$r_e = b_{eff} \quad (19)$$
$$c_e = (1/k_{eff})$$

The *force-current analogy* could have also been used to go between electrical and mechanical systems, where correspondence is made between force and current, velocity and voltage, mass and capacitance, etc.

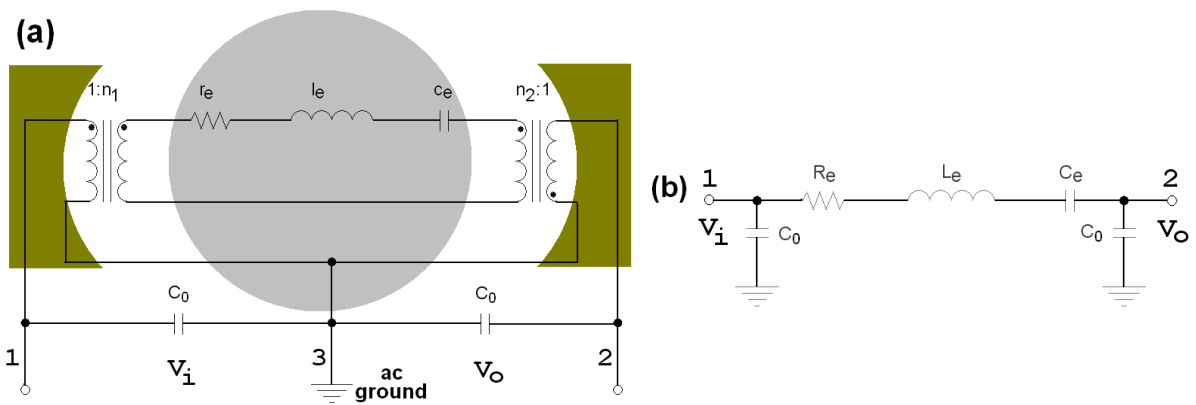

**Fig. 7.** (a) An equivalent electrical model for the resonator. The disk and I/O electrodes are shown in the background for ease of understanding. (b) Simplified model by using reflection through the transformers.

Since micromechanical resonators are used in electronic circuits, it is necessary to convert electrical energy to mechanical energy at the input, and the reverse at the output. This is exactly what is accomplished by the two electromechanical transducers present at the input and output ports of the resonator. The schematic and equivalent circuit representations of such a transverse (or, parallel-plate) electrostatic transducer are given in Fig. 8. The transfer equations relating the effort-flow variables at the electrical port to those at the mechanical port for this transducer can be expressed as follows (Tilmans 1996):

$$\begin{bmatrix} v \\ i \end{bmatrix} = \begin{bmatrix} \frac{1}{n} & \frac{n}{j\omega C_0} \\ \frac{j\omega C_0}{n} & 0 \end{bmatrix} \begin{bmatrix} F \\ u \end{bmatrix} = \begin{bmatrix} 1 & 0 \\ j\omega C_0 & 1 \end{bmatrix} \begin{bmatrix} \frac{1}{n} & 0 \\ 0 & -n \end{bmatrix} \begin{bmatrix} 1 & \frac{n^2}{j\omega C_0} \\ 0 & 1 \end{bmatrix} \begin{bmatrix} F \\ u \end{bmatrix} \qquad (20)$$

Here voltage (v) and force (F) are the effort-variables, and current (i) and velocity (u) are the flow-variables. Also, $C_0$ denotes the static capacitance between the parallel plates and n is called the *transduction factor* or *electromechanical-coupling coefficient* which is the transformer-ratio of the transducer. In the right-hand-side of (20), the matrices represent the electrical admittance, the transducer portion, and the mechanical impedance respectively. Each of these constituent transfer matrices can be represented by an equivalent circuit, with the overall equivalent electrical model of the transducer being represented by a cascade-connection of these sections as depicted in Fig. 8(b).

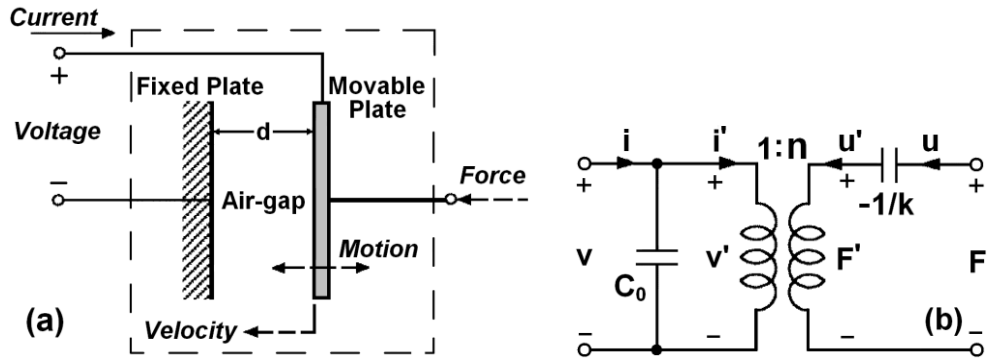

**Fig. 8.** Schematic diagram (a) and electrical equivalent circuit (b) of a transverse electrostatic transducer.

Hence, in the disk-resonator equivalent circuit of Fig. 7(a), the electromechanical transduction at the two ports is modeled by the transformers with turns-ratio $n_k$ given in (21).

$$n_k = V_{dc} \frac{\partial C_k}{\partial r} = V_{dc} \frac{\partial}{\partial r}\left(\frac{\varepsilon A_k}{d_0 - r}\right) \approx V_{dc}\left(\frac{\varepsilon A_k}{d_0^2}\right) \quad \text{(for, } r << d_0) \qquad (21)$$

Here $C_k$ denotes the capacitance between the disk and the $k^{th}$ electrode (k = 1 or 2), $A_k$ is the coupling area given by ($\phi_k R t$), $\phi_k$ is the angular overlap of the electrode with the disk, and r is the lateral displacement. By means of reflection of the motional elements ($r_e$, $l_e$ and $c_e$) in the mechanical domain through the transformers to the electrical domain at the I/O-ports, a simpler electrical equivalent circuit of the resonator can be derived as shown in Fig. 7(b). The corresponding element values are (assuming symmetrical electrodes i.e., $n_1 = n_2 = n$):

$$L_e = \left(\frac{l_e}{n^2}\right)$$
$$C_e = (n^2 c_e) \quad (22)$$
$$R_e = \left(\frac{r_e}{n^2}\right)$$

Actually $C_e$ is given by $n^2/(k_{eff} - 2k)$ where $k$ is the effective spring constant due to transduction. But, the approximation of (22) is preferred as the value of the mechanical stiffness ($k_{eff}$) is many times ($\sim 10^6$) larger than the value of electrical stiffness in general (Wang et al. 2004a). The value of the motional-resistance $R_e$ realized in a particular resonator design is very crucial and should be low enough for proper matching with conventional RF stages which typically have an impedance of 50 Ω. Using (13), (16), (18) and (19–22), the following expression of $R_e$ can be derived for the first radial-bulk mode:

$$R_e = \left(\frac{\omega_0 m_{eff}}{Q V_{dc}^2}\right)\left(\frac{d_0^4}{(\varepsilon_0 \phi_1 Rt)^2}\right) = \left(\frac{1.18 \times 10^{29}}{Q V_{dc}^2}\right)\left(\frac{d_0^4}{Rt}\right) \quad (23)$$

This calculation is for a polysilicon disk with the assumptions of $\phi_1 = \phi_2 \approx \pi$ and $\varepsilon = \varepsilon_0$. Thus, the most effective way to reduce the motional resistance is to decrease the transduction gap $d_0$ to deep-submicron level and use thicker layers to make the disk. But, the process to be used for fabricating the resonators may act as a hindrance towards this end. For example, the minimum allowable space between features specified in the design rules for the process constraints the minimum value of the air-gap achievable. A large value of dc-bias can also reduce $R_e$, but it may not be feasible always.

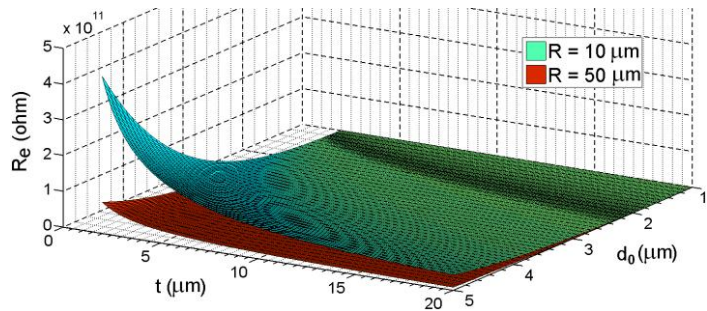

**Fig. 9.** Plot showing variation of series motional resistance ($R_e$) with disk-to-electrode gap ($d_0$) and disk thickness (t) for two values of disk radius (R) for a fixed assumed Q of 10,000 and $V_{dc}$ of 30 V.

The nature of variation of the value of motional resistance with the gap, disk-thickness and disk-radius is illustrated in Fig. 9.

## 5.5 Model-to-Hardware Correlation

The theoretical background and modeling predictions discussed in the previous sections can be validated by means of proper characterization of the resonator devices. Different techniques can be used for this purpose; like one-port measurement, two-port measurement, mixing measurement, RF/LO measurement etc. (Clark et al. 2005); with each having their own advantages and disadvantages. A comparison among the measured and predicted performances for a polysilicon disk resonator with R = 18 µm, t = 2.1 µm and $d_0$ = 87 nm, reported by Wang et al. (2004a) is repeated here in Table 7; which shows very good agreement. Analytical values can also be verified using finite-element method (FEM) based simulation (Basu and Bhattacharyya 2011). Thus, the modal resonance frequencies, corresponding mode-shapes, frequency response characteristics, etc. can be obtained form such simulations.

**Table 7.** Comparison of measured and predicted values for a disk resonator (Wang et al. 2004a)

| Parameter | 1st resonance mode (analytical) | 1st resonance mode (measured) | Unit |
|---|---|---|---|
| Frequency, $f_0$ | 152.4 | 151.3 | MHz |
| Quality factor, Q | – | 12,289 (vacuum) 9,316 (air) | – |
| dc-bias, $V_{dc}$ | – | 6 | V |
| Mass, $m_{eff}$ | $3.83 \times 10^{-12}$ | – | kg |
| Stiffness, $k_{eff}$ | 3.52 | – | MN/m |
| Damping factor, $b_{eff}$ | $3.01 \times 10^{-7}$ | – | kg/s |
| Series resistance, $R_e$ | 479.526 | 479.667 | kΩ |
| Series inductance, $L_e$ | 6.20 | 6.19 | H |
| Series capacitance, $C_e$ | $1.78 \times 10^{-19}$ | $1.79 \times 10^{-19}$ | F |

## 6. Energy Loss and the Quality Factor

The mechanical quality factor (Q) of a microresonator is a dimensionless parameter that models the losses in the vibrating structure. It is defined in the following manner:

$$Q = 2\pi \left( \frac{\text{Energy stored}}{\text{Energy dissipated per cycle}} \right) \quad (24)$$

It is also defined as the ratio of the resonant frequency to the width of the resonant peak:

$$Q = \frac{f_0}{\Delta f} \quad (25)$$

Here $f_0$ is the center-frequency of the peak, usually referred to as the damped resonant frequency, and $\Delta f$ is the half-power or 3dB bandwidth around the measured peak. The quality factor is also related with the damping ratio ($\zeta$) by:

$$Q = (1/2\zeta) \quad (26)$$

Quality factor is proportional to the decay time. So, higher frequency stability and accuracy can be obtained from resonators with higher Q. Energy loss in resonators can occur in the form of many distinct mechanisms, such as air damping, support loss, thermoelastic damping, internal friction etc. (Brotz 2004; Merono 2007). A brief discussion on such sources of damping in MEMS resonators is presented here.

- When a body travels through a fluid, it collides with the molecules of the fluid and transmits some of its energy to those molecules. This contributes to two components of damping. One is due to the friction of the vibrating resonator structure with the fluid, and another is due to the squeezing of the fluid present between the moving resonator and the closely placed electrodes. These loses are denoted by *air-damping* and *squeeze-film damping* respectively. The energy-loss due to squeeze-film damping dominates when it exists, as in the case of capacitively driven and sensed microresonators.

- Another dominant source of loses is the coupling of energy from the resonator through its supporting anchors to the surrounding structure. This leakage can be minimized by designing a balanced resonator e.g., by supporting the resonator at its nodal points. This type of dissipation is called *anchor-loss* or, *structural damping*.

- *Surface effects* can also act as loss mechanisms in MEMS resonators, although this hasn't received much theoretical attention till now. One such loss may be due to *surface mechanical dissipation* originating from a thin layer of adsorbates on the structural layer surface, or due to *plasma damage* produced on the surface during Reactive Ion Etching (RIE). It has been experimentally demonstrated that the Q can be enhanced by a rapid thermal annealing (RTA) at 600-800ºC for about 30 s after the fabrication of the resonators.

- *Thermoelastic damping* (TED) occurs when certain regions of the resonator structure are experiencing compression while others are in expansion, leading to stress gradients. The laws of thermodynamics predict that compressed materials tend to heat while materials in expansion tend to cool. The thermal gradient produced between these two regions induces irreversible energy loss due to flow of heat. This cyclic energy loss is a function of the frequency of vibration and is significant in resonant beam structures. However, as for a radial-contour mode disk resonator all areas are entirely in expansion or contraction, thus this heat flow is minimized.
- Damping may also be due to internal material parameters, such as crystal structure, lattice defects etc. by which vibrations can get attenuated. This mechanism is named as *internal friction*.

Since energy-loss sources add directly for total energy-loss, individual Q's should be added reciprocally for yielding the total quality factor:

$$\frac{1}{Q} = \frac{1}{Q_{Air-damping}} + \frac{1}{Q_{Air-squeezing}} + \frac{1}{Q_{Anchor-loss}} + \frac{1}{Q_{TED}} + \frac{1}{Q_{surface}} + \frac{1}{Q_{internal}} \quad (27)$$

For MEMS resonators operating in air, Q is limited by the air damping and squeezing; while in a vacuum ambient, the main losses are due to the supporting anchors, with TED and surface effects playing a minor role. Also, the Q of high frequency resonators is much less susceptible to ambient pressure than the lower frequency flexural beam resonators. Thus, for bulk-mode disk resonators, Q gets limited mainly by anchor dissipation rather than by viscous damping and other losses (Wang et al. 2004a).

## 7. Potential Applications of MEMS Resonators

Instead of interconnecting at the board level, integration of the discrete components (like crystal oscillator reference, band-select filter, image-reject filter, channel-select filter etc.) with the large-scale integrated circuits (like amplifiers, mixers, oscillators etc.) producing a single IC chip will lead to dramatic decrease in the size of telecommunication transceivers. The corresponding reduction in off-chip macro-scale component count will also lead to lower production, packaging and handling costs, and the elimination of board-level interconnect parasitics; hence, improving the performance and reliability considerably for high-frequency applications. But, the lack of high Q on-chip passive components in even the most modern IC processes poses a serious bottleneck towards this end. In this context, high-Q on-chip micromechanical resonators have emerged as the key element in the realization of filters and oscillators, as their Q determines the insertion loss and phase noise, respectively.

## 7.1 Oscillators Based on MEMS Resonators

An oscillator is defined as a system that generates a periodic signal from a constant energy input and a single timing reference. So it is a circuit comprising of a timing element and a gain element; with the output frequency being ideally defined entirely by the timing element. Leeson's equation indicates that the stability of an oscillator, as measured by its phase noise, is inversely proportional to the Q of its frequency-setting tank element (Leeson 1966). MEMS resonators offer the potential for very high Q's and thus are ideal for use as the timing element in oscillator implementations. An electronic amplifier is typically used as the gain element. The whole system is designed to operate under the conditions of positive feedback. The loop gain of the oscillator must be designed to be larger than unity and the phase of the gain should be equal to zero at the output frequency of the oscillator. Kaajakari et al. (2004) have demonstrated a 13.1 MHz square extensional mode resonator (with Q of 130,000) based oscillator having a single-sideband phase noise of –138 dBc/Hz at 1 kHz offset from the carrier and a noise floor of –150 dBc/Hz which satisfies the GSM specifications (typically –130 dBc/Hz at 1 kHz) usually achieved only with macroscale quartz crystals. The oscillator is based on a discrete amplifier connected in series feedback configuration with the resonator inside a vacuum chamber.

Researchers have reported the usage of either a sustaining *transresistance amplifier*, or a *Pierce amplifier* for making oscillators, with the MEMS-based resonator acting as a frequency selective component present in the feedback path (Vittoz et al. 1988; Wong and Palaniapan 2009). Monolithic integration of CMOS and MEMS has been realized in such efforts. An overview of the research on these two types of oscillators is provided in the following two subsections.

### *7.1.1 Transresistance Oscillator*
A transresistance amplifier outputs a voltage proportional to its input current, and is often referred to as a *transimpedance amplifier* (TIA). This can also be described as a *current-controlled voltage source* (CCVS). Since, microelectromechanical resonators have a voltage-in and current-out transfer function, so the sustaining amplifier has to be designed as a TIA. The first fully-integrated CMOS-MEMS oscillator was demonstrated by Nguyen and Howe (1993) utilizing a comb-drive resonator fabricated along side with sustaining TIA CMOS electronics on a single chip. Fig. 10 presents the top view SEM of the prototype of such an oscillator. But, the oscillation frequency was only about 16.5 kHz and not sufficient for wireless communication applications. Also, measurements for phase noise performance were not reported for this oscillator. Later, using a wine-glass mode disk resonator, Lin et al. (2004a) realized a 60 MHz reference oscillator yielding a phase noise density of −100dBc/Hz at 1 kHz offset from the carrier, and −130dBc/Hz at far-from-carrier offsets. Recent studies on micromechanical oscillators have shown desirable phase noise performance, but have also revealed additional sources of phase noise related to nonlinear amplitude perturbations caused by 1/f noise (Roessig et al. 1997). A reference oscillator based on a composite array of nine 62 MHz wine-glass mode disk resonators, has been demonstrated with an impressive normalized phase noise performance (Lin et al. 2005). Thus, going for an array-structure instead of a stand-alone resonator helps in achieving a tank element that handles significantly higher power, while still retaining a very high Q of 118,900.

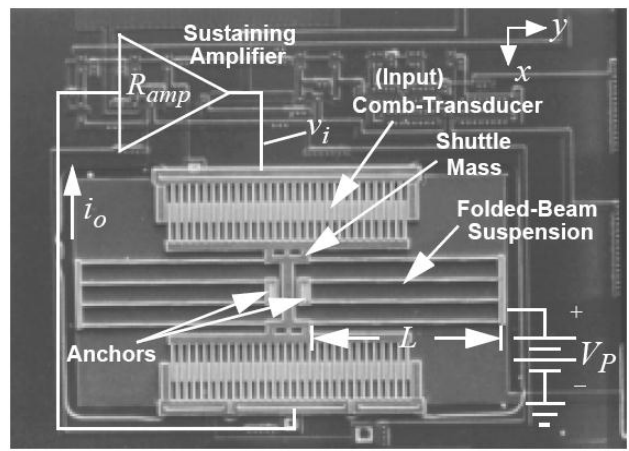

**Fig. 10.** SEM of the 16.5 kHz comb-drive microresonator based CMOS oscillator with schematics explicitly depicting the circuit topology. [Reprinted with permission from (Nguyen 2007). Copyright (2007) IEEE.]

*7.1.2 Pierce Oscillator*

The schematic of a Pierce oscillator is given below. It consists of an inverting amplifier, two capacitors ($C_1$ and $C_2$), and the resonator (Fig. 11). The damping due to the motional resistance gets compensated here by the negative resistance of the transconductance amplifier leading to oscillations. The Pierce-topology is actually favorable than the transresistance oscillators which have too many active components and thus, may lead to unacceptable power consumption and phase-noise levels. Roessig et al. (1998) have demonstrated a 1 MHz CMOS-MEMS Pierce oscillator with a far-carrier phase noise of −89 dBc/Hz. Subsequently,

Lee et al. (2001) have demonstrated a 9.75 MHz resonator using a CC-beam microresonator tank, with a phase noise performance of –80 dBc/Hz at a 1 kHz frequency spacing. Mattila et al. (2002) have reported a 14 MHz oscillator based on a CC-beam as the timing element with a far-carrier noise floor of –120 dBc/Hz and a near-carrier noise of −105 dBc/Hz at 1 kHz offset. Recently, a 1.05 GHz Pierce oscillator based on a piezoelectric MEMS resonator (Q of 2,200) wire-bonded to the CMOS IC chip has been demonstrated (Zuo et al. 2010). It has shown a phase noise level of –81 dBc/Hz at 1 kHz offset frequency, and a phase noise floor of –146 dBc/Hz, with a dc power consumption of 3.5 mW.

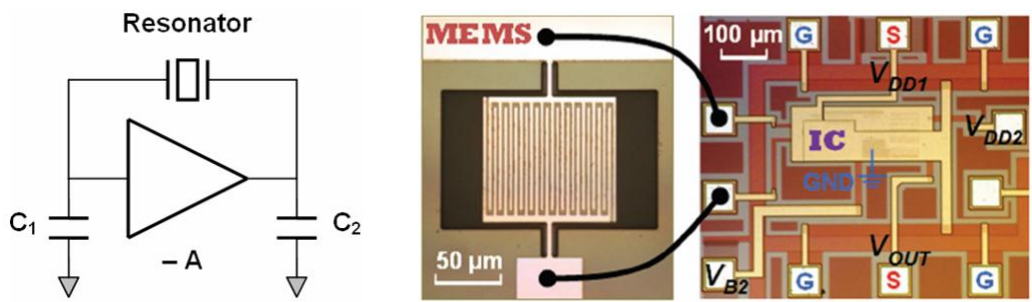

**Fig. 11.** Circuit schematic of a Pierce oscillator circuit. The micrograph [reprinted with permission from (Zuo et al. 2010), copyright (2010) IEEE] shows a Pierce oscillator with the MEMS resonator wire-bonded to the amplifier IC.

## 7.2 Filters Based on MEMS Resonators

A single resonator is a first-order resonant system providing limited frequency selectivity. Thus, for realizing bandpass filters (BPF) with higher selectivity, higher order resonant systems comprising of a number of coupled resonators are required. Two approaches that can be used for implementing a BPF using micromechanical resonators are the usage of electrical and mechanical coupling.

### 7.2.1 MEMS BPF Using Electrical Coupling

In this approach, the individual resonators are cascaded with shunt capacitors to ground in between every two adjacent resonators. The coupling capacitors interact with the equivalent RLC tanks of the microresonators resulting in several resonance modes and consequently a multiple order bandpass frequency response for the system. Fig. 12 depicts such a BPF realized using single-crystal silicon flexural mode CC-beam resonators (Pourkamali 2004; Pourkamali and Ayazi 2005). This has demonstrated a filter quality factor of 1500 at an operating frequency of 810 kHz, bandwidth of 540 Hz, stopband rejection of 28 dB, and 20dB shape-factor of 2.9.

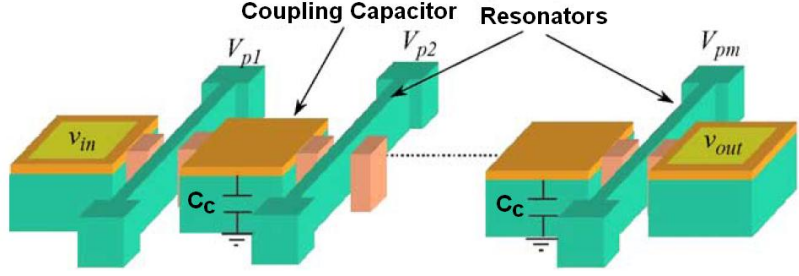

**Fig. 12.** Schematic diagram of a capacitively coupled MEMS bandpass filter. [Reprinted with permission from (Pourkamali and Ayazi 2005). Copyright (2005) Elsevier.]

For a two-resonator (second order) capacitively coupled resonant system, a new pair of conjugate poles appears in the transfer function of the system and hence, a new resonance mode at a frequency $f_1$ given by:

$$f_1 = f_0 \sqrt{\frac{1 + \pi f_0 C_C R_e Q}{\pi f_0 C_C R_e Q}} \qquad (28)$$

Here $f_0$ is the center frequency of each resonator, $C_C$ is the shunt coupling capacitor, and Q and $R_e$ denote the quality factor and motional resistance of an individual resonator respectively. The creation of the extra resonance mode can be explained using the electrical equivalent circuit of the second-order capacitively coupled resonant system which is given in Fig. 13 along with the corresponding frequency response. The input and output capacitances have been neglected for simplicity.

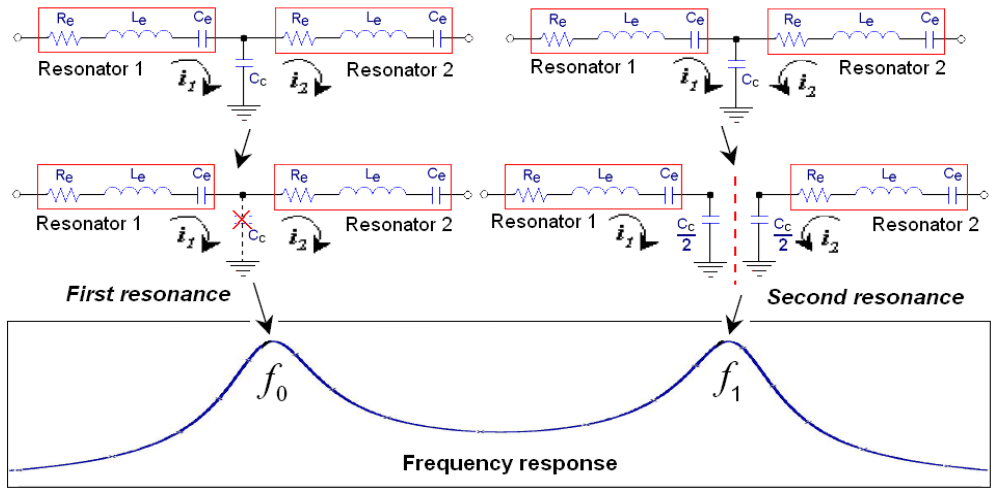

**Fig. 13.** Equivalent electrical circuits of a second order capacitively coupled BPF in the first and second resonance modes and the overall frequency response.

Now, the first resonance occurs at the mechanical resonant frequency of the individual resonators ($f_0$) when they resonate in phase and the coupling capacitor has negligible contribution (while $C_c$ is being charged by the first resonator, the other resonator is discharging it). However, in the second resonance mode, the two resonators vibrate out of phase, and hence the coupling capacitor is either charged or discharged at the same time by both the resonators. Utilizing the symmetry of the equivalent circuit, it can be split into two half circuits, each consisting of one resonator and a series capacitor $C_c/2$ to ground. This coupling capacitor ($C_c/2$) in series with the motional capacitance of the resonator ($C_e$) reduces the total capacitance of the equivalent RLC tank producing the second resonance mode at a higher frequency $f_1$.

Another alternative electrical coupling technique is the active cascading of resonators using amplifying interface circuits in between contiguous resonators to eliminate the loading effects. Thus, there occurs a multiplication of the individual transfer functions and an overall higher order transfer function for the system. Researchers have also shown that this technique leads to an enhancement of the equivalent Q, smaller shape-factor (hence, higher selectivity), and sharper roll-off for the overall system. A BPF of wider bandwidth can be achieved by introducing a slight mismatch between the center frequencies of the cascaded resonators (Pourkamali and Ayazi 2005). Also, the characteristics of such filters can be widely tuned by changing the dc polarization voltages ($V_p$) used.

### *7.2.2 MEMS BPF Using Mechanical Coupling*

Another popular technique for synthesizing higher order filters using MEMS resonators is their mechanical (or, acoustic) coupling. The theory of vibrations suggests that the number of resonance modes of a mechanical system is equal to the number of degrees of freedom (Johnson 1983). Thus, if a number of microelectromechanical resonators are physically connected to each other by means of compliant coupling elements (springs); the ensuing structure will have different resonance modes, with the number of modes being equal to the number of coupled resonators. The most widespread coupling method is the attachment of one resonator to another by means of one or more thin coupling beams. The mode of vibration of the individual resonators determines the vibration mode for the coupling beams which can be either *extensional*, or *torsional*, or *flexural* (Johnson 1983). Thus for example, for a BPF realized by means of radial-contour or wine-glass mode disks vibrating in the lateral direction, the required coupler beam modes should be extensional in type.

For the case where the coupled resonators have equal vibration frequencies, the first resonance mode of the coupled array will occur at the same natural frequency as that of the individual resonators, and the rest of the resonance modes will occur at close but slightly higher frequencies. The frequency difference between these resonance modes is determined by the ratio of the stiffness of the resonator ($k_r$) to the stiffness of the coupling element ($k_{sij}$). As illustrated in Fig. 14(a), two CC-beam resonators can be coupled by a soft flexural-mode mechanical spring attached to each resonator at low-velocity locations to constitute a second-order BPF, and two transducers are required to interface this filter with the electrical counterpart (Bannon et al. 1996; Bannon et al. 2000). The two distinctive mode-shapes of this system are also shown in Fig. 14(c).

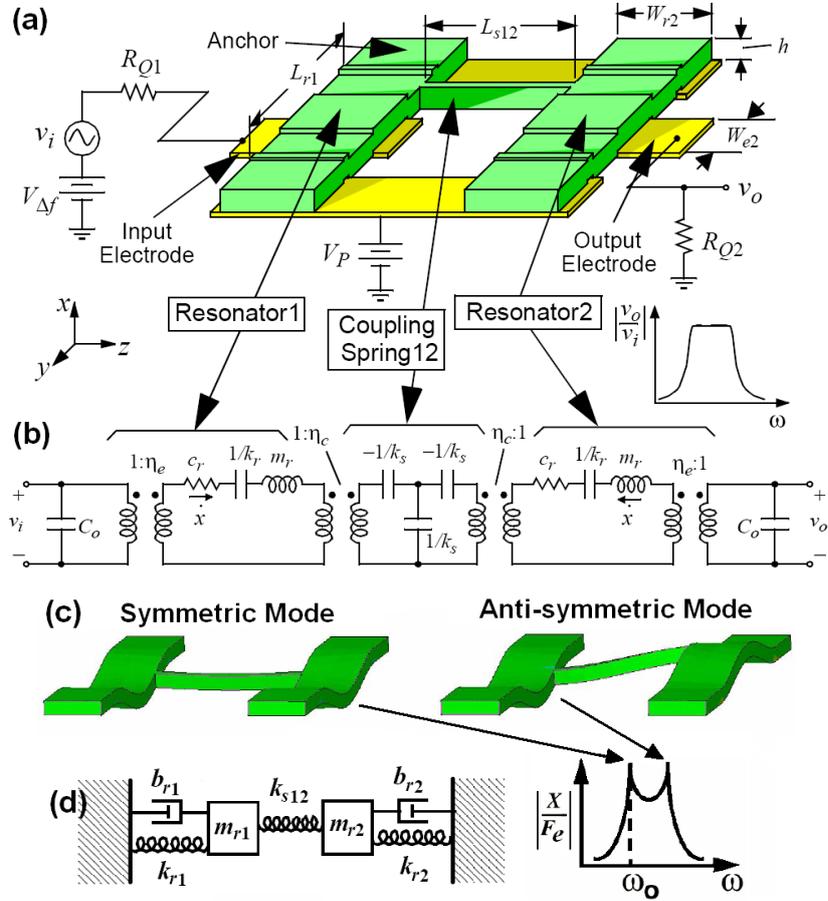

**Fig. 14.** (a) Schematic view of a two-resonator micromechanical filter. (b) The equivalent electrical circuit for the filter. (c) Filter mode shapes and their correspondence to specific peaks in the (displacement/force) frequency characteristic. (d) The equivalent mechanical circuit of the BPF. [Reprinted with permission from (Bannon et al. 2000). Copyright (2000) IEEE]

In the first resonance mode, which is of the lowest frequency, all the adjacent resonators vibrate in phase. This is due to the fact that the coupling springs are not deformed here, and hence, has no effect on the vibration of the resonators and the resonance frequency will be the same as the resonance frequency of a single resonator (assuming negligible mass of the coupling beams). In the second resonance mode, all the resonators vibrate out of phase and the equivalent stiffness of each individual resonator increases by twice the stiffness of the coupling spring causing a shift in the resonance frequency as shown in the figure. The creation of subsequent resonance modes can be similarly explained for third and higher order filters. Having several resonance peaks at close frequencies provides sharper transition from passband to the stopband, making the filter characteristic closer to that for an ideal BPF.

The equivalent electrical and mechanical models for the second-order filter are provided in Fig. 14(b) and (d). The termination resistors $R_{Qn}$ should be properly chosen to flatten the jagged passband of Fig. 14(c) to achieve that shown in Fig. 14(a). The frequency of the second resonance peak ($f_1$) is related to that of the first ($f_0$) by the following relation (Pourkamali 2004):

$$f_1 = f_0 \sqrt{1 + \frac{2k_{s12}}{k_r}} \tag{29}$$

Using this geometry, Wong et al. (1999) have demonstrated two-pole BPFs with center frequencies from 50–68 MHz and percent bandwidths less than 2%, all with insertion losses less than 9 dB. A BPF using three identical folded-beam comb-drive resonators coupled by flexural-mode beams was first reported by Wang and Nguyen (1999), having a bandwidth of 403 Hz at a center-frequency of 340 kHz, stopband rejection of 64 dB and insertion loss less than 0.6 dB. Demirci and Nguyen (2006) have used strong mechanical coupling of transverse-mode square resonators at their corners to form an array, where all the resonators vibrate at a single mode frequency with other modes being suppressed by electrode phasing. Thus, it allows matching to a much lower 50 Ω impedance. By using such mechanically coupled flexural-mode square resonator arrays as *composite resonators*, a two-resonator 68.1 MHz filter has been demonstrated allowing its matching to 50 Ω termination impedances (Demirci and Nguyen 2005). An insertion loss less than 2.7 dB for a 190 kHz passband width and a stopband rejection of 25 dB have been exhibited here. MEMS filters up to the 20th order have been demonstrated using mechanically coupled drumhead resonators with electrical actuation and optical sensing schemes at center frequencies of a few MHz (Greywall and Busch 2002). A high-Q disk array composite filter using two four-disk array composite resonators coupled by extensional-mode beams has been reported with a frequency of 156 MHz and bandwidth of 201 kHz, providing a stopband rejection of 22 dB and insertion loss less than 2 dB (Li et al. 2006).

## 8. MEMS-CMOS Integration

Two distinct techniques are available for integrating the MEMS device with the IC: system-in-package (SiP), and system-on-chip (SoC). In the *system-in-package* or *hybrid* or *multi-chip* or *wafer-bonding* approaches, the CMOS and MEMS chips are fabricated separately on different substrates and bonded at the wafer level. This is the more dominant approach going into MEMS products and manufacturing at present due to modularity, IP issues, yield, development cycle and time to market challenges as compared to the monolithic approach. Independent optimization of the MEMS and CMOS technologies is also possible. Researchers have used this technique for testing the prototypes of their MEMS resonator based RF devices (Mattila et al. 2002; Sundaresan et al. 2005; Zuo et al. 2010). But, the drawback of this method is the generation of parasitics due to interconnections and bond-pads, which may impede the system's performance especially at higher frequencies. Also, the assembly and packaging cost and complexity is in general higher in this process.

Fabrication options for the *system-on-chip* concept i.e., monolithic integration of MEMS devices with CMOS circuit can be categorized into pre, intra, and post-CMOS approaches; with each having their own advantages and disadvantages. This offers the prospect of lower cost and parasitics, and higher integration density and speed than in the SiP process; but with an increase in the complexity of the overall design. In *pre-CMOS* or *MEMS-first* integration, the MEMS structures are formed prior to the regular CMOS process sequence. SOI wafers are normally favored for fabricating the microstructures before sending it to the IC foundry. But the approach is not much feasible as most CMOS foundries would not work with the pre-processed wafers in order not to compromise their process yield. The surface planarization required for the subsequent CMOS processing can also act as a restraint. A MEMS-first wafer-level encapsulated microresonator has been demonstrated by Candler et al. (2006), where the resonator can be released and encapsulated (at a pressure < 1 Pa) prior to the start of CMOS processing steps.

In the *intermediate-CMOS* approach, some additional thin film deposition or micromachining steps are performed in between the regular CMOS steps. The mixing of the MEMS and CMOS fabrication steps has only been successfully realized by Analog Devices, Inc. (Nguyen 2009). But, a major drawback of this process for fabricating RF systems is the requirement of large channel-lengths (for enduring the high thermal budget for fabrication of MEMS) that can adversely affect the cut-off frequency of the associated circuit.

In the *post-CMOS* or *MEMS-last* micromachining approach, MEMS devices are processed on top of the CMOS substrate. Two separate strategies can be utilized for this: (a) building the mechanical structures on top of the CMOS without disturbing any of the CMOS layers; and (b) building the microstructures by machining the CMOS substrate wafer itself after completion of the regular CMOS process sequence. A key advantage of this is the modular approach, and the fabrication can be completely outsourced to a dedicated MEMS foundry without introducing any changes in the standard CMOS process steps. However, the problem with post-CMOS procedure is that the assembling of MEMS components requires high temperatures that can damage the copper or aluminum metal layers, and the low-k dielectrics used in the underlying electronic devices. This limits the MEMS processing temperatures to about 400 °C. The etchants used while post-processing can also affect the passivation layer of CMOS. Also, the wafers should be of equal dimensions in the MEMS foundry as for the CMOS processing which is seldom the case. An example of a post-CMOS integrated microresonator has been reported by Takeuchi et al. (2004) using the damascene process. MEMS resonators fabricated using nickel, SiGe, etc. also seem to be favorable due to a much lower thermal process budget (Huang 2008; Quevy et al. 2006). Huikai et al. (2002) have fabricated resonators and other high aspect-ratio MEMS devices co-fabricated with conventional CMOS process using the post-CMOS route. MEMS has been monolithically integrated with CMOS amplifier circuitry using the post-CMOS approach for realizing a mixer-filter for down-conversion of RF signals to an intermodulation frequency by Fedder (2005). An extensive review of the various CMOS-MEMS fabrication technologies is beyond the scope of this paper, and further details can be found in the following references: Brand (2006); Brand and Fedder (2005); Fedder et al. (2008).

## 9. Future Outlook and Conclusion

A recently conducted analysis on emerging MEMS products has stipulated that MEMS oscillators will take a significant share of the timing market by 2015 with a compound annual growth rate of 80% from 2010 to 2015; with the forecasted value reaching USD 644.9 million in 2015 from just USD 7.9 million in 2009 (Yole 2010). The aggressive growth of MEMS research, development, and marketing has even provided a shot in the arm for the quartz crystal and crystal oscillator industry which has made major progresses in the past decade with regard to miniaturization, performance and reliability improvement, cost reduction, and new application development (Lam 2008).

The smart transceivers of tomorrow would demand the integration of micromechanical and microelectronic systems leading to consolidation of all the RF/analog/digital components onto a highly miniaturized single chip, commonly referred to as RF system-on-a-chip (RF-SoC) (Brand and Fedder 2005). In particular, high-Q micromechanical circuits can enable paradigm-shifting transceiver architectures that trade power for selectivity (i.e., Q), hence having the potential for significant power savings and multiband reconfigurability (Nguyen 2007). But the most formidable challenge towards this is the aspect of the technology used for fabrication. Firstly, the design rules for the process should permit narrow enough feature size and spacing so that the desirable electrode gap (for reduced equivalent motional resistance) and anchor dimensions (for reduced anchor-losses) can be achieved. Also, the thickness of the structural layer for MEMS should be large enough for increasing the transduction area and the power handling capacity of the resonator. Secondly, considerable research should be dedicated towards the development of appropriate CMOS-MEMS processes for addressing the various challenges of MEMS integration. Also, the material properties should be adequately known as the physics are often different due to non-traditional size scales. There should be a very tight control on the structural dimensions (with a high aspect-ratio) and the materials used for accurate frequency reproducibility.

Another major hurdle for commercialization of MEMS resonators is the frequency-stability against time and temperature. A notable solution towards the achievement of commercial level long-term stability for silicon resonators can be their packaging in a controlled environment by *epi-seal* encapsulation process (which is a wafer-scale thin-film encapsulation using epitaxial silicon deposition), minimizing most of the potential aging mechanisms like contamination-decontamination process, pressure change, and diffusion effects (Kim et al. 2007a). For competing against the mature quartz technologies, packaging of the MEMS resonators is surely a vital aspect which is much more complex than for ICs. The packaging should leverage MEMS strengths, namely small size, potential of CMOS compatibility, low cost etc., while preventing frequency drift due to packaging contamination (Esashi 2008; Partridge et al. 2005). A good review of MEMS packaging techniques can be found in (Tummala and Swaminathan 2008). But, unlike long-tem stability, the frequency stability versus temperature remains one of the fundamental challenges in the design and fabrication of micromechanical resonators. This is mainly due to the negative Young's modulus temperature coefficient of silicon that leads to a decrease in the resonator center frequency with temperature. Generally, MEMS resonators made of typical materials including silicon, exhibit an inherent drift over temperature of approximately –20 ppm/°C. To attain the frequency stability claimed by the MEMS oscillator suppliers, temperature compensation becomes indispensable. One approach is the design of complex ASICs to compensate for the frequency inaccuracies due to environmental temperature changes with a feedback loop, inevitably adding size, cost, power consumption, and phase noise (Cioffi et al. 2010; Ho et al 2006; Hopcroft et al. 2006). Another strategy that doesn't require additional power or increase in size or control circuit is the introduction of one or more number of materials, other than silicon, and utilizing the difference in their mechanical properties (Hsu et al. 2000; Hsu and Nguyen 2002; Melamud et al. 2005). For example, researchers have used the thermal expansion mismatch between silicon and the compensating material (like gold or aluminum) towards this approach. But, the intrinsic mechanical instability of such metals, like fatigue and hysteresis produce long-term stability issues with this technique. Some success have been recorded to temperature-compensate the silicon MEMS resonator with the deposition of silicon dioxide that becomes stiffer as temperature increases exhibiting up to 40-times improvement in resonant frequency stability (Kim et al. 2007b). However, the lossy $SiO_2$ tends to de-Q the MEMS resonator quickly. An initial frequency accuracy of about 1.5 ppm is also hard to achieve with such resonators. Therefore, maintenance of the temperature steadiness of resonant frequency deserves considerable attention. With regard to the dynamic range of such devices, the non-linearity in the drive and sense determines the upper-bound (Mestrom et al. 2008), and noise determines the low end. So, proper understanding of these aspects is also crucial. Further, the quest for innovative geometries of MEMS resonators should be carried forward so that the best performance can be obtained out of them (Basu and Bhattacharyya 2011). Thus, a *vertical resonator* instead of the lateral transduction mechanism can provide us with reduced equivalent resistance (Pomarico et al. 2010). Advanced computer-aided design (CAD) tools should be developed which can allow the simulation of integrated MEMS and CMOS components on the same platform; and can automatically generate micromechanical circuits for a given specification. It can be concluded that the evolution of vibrating RF MEMS is bound to accomplish the targets in the near future through an active collaboration between the industry and the academic R&D establishments.

## Acknowledgements

The authors would like to express their deep gratitude to National Programme on Micro and Smart Systems (NPMASS), Govt. of India for extending the infrastructure of MEMS Design Laboratory at Indian Institute of Technology Kharagpur; and thank the anonymous reviewers of this journal for their valuable comments and suggestions on the improvement of this article.

## References

Abdelmoneum MA, Demirci MU, Nguyen CTC (2003) Stemless wine-glass-mode disk micromechanical resonators. In: Proceedings of the 16th IEEE International Conference on Micro Electro Mechanical Systems, Kyoto, Japan, Jan 2003, pp. 698–701

Bannon FD, Clark JR, Nguyen CTC (1996) High frequency microelectromechanical IF filters. In: Technical Digest of IEEE International Electron Devices Meeting, San Francisco, CA, Dec 1996, pp. 773–776

Bannon FD, Clark JR, Nguyen CTC (2000) High-Q HF microelectromechanical filters. IEEE J Solid-State Circ 35:512–526

Basu J, Bhattacharyya TK (2011) Comparative analysis of a variety of high-Q capacitively transduced bulk-mode microelectromechanical resonator geometries. Microsyst Technol 17(8):1361–1371

Bhave SA, Di G, Maboudian R, Howe RT (2005) Fully-differential poly-SiC Lame mode resonator and checkerboard filter. In: Proceedings of the 18th IEEE International Conference on Micro Electro Mechanical Systems, Miami, Florida, Jan–Feb 2005, pp. 223–226

Brand O (2006) Microsensor integration into systems-on-chip. Proc IEEE 94(6):1160–1176

Brand O, Fedder GK (2005) Advanced micro & nano systems, volume 2: CMOS-MEMS. Wiley-VCH, Weinheim

Brotz J (2004) Damping in CMOS-MEMS resonators. Master's thesis, Carnegie Mellon University

Candler RN, Hopcroft M, Kim B, Park WT, Melamud R, Agarwal M, Yama G, Partridge A, Lutz M, Kenny TW (2006) Long-term and accelerated life testing of a novel single-wafer vacuum encapsulation for MEMS resonators. J Microelectromech Syst 15(6):1446–1456

Chandorkar SA, Agarwal M, Melamud R, Candler RN, Goodson KE, Kenny TW (2008) Limits of quality factor in bulk-mode micromechanical resonators. In: Proceedings of the 21st IEEE International Conference on MicroElectroMechanical Systems, Tucson, Arizona, Jan 2008, pp. 74–77

Cioffi KR, Hsu WT (2005) 32 KHz MEMS-based oscillator for low-power applications. In: Proceedings of the 2005 IEEE International Frequency Control Symposium and Exposition, Vancouver, BC, Aug 2005, pp. 551–558

Cioffi KR, Simoneau M, Lacroix D, Hsu WT (2010) Counter-based resonator frequency compensation. US Patent 7679466

Clark JR, Hsu WT, Abdelmoneum MA, Nguyen CTC (2005) High-Q UHF micromechanical radial-contour mode disk resonators. J Microelectromech Syst 14 (6):1298–1310

De Los Santos HJ, Fischer G, Tilmans HAC, van Beek JTM (2004) RF MEMS for ubiquitous wireless connectivity: Part II–Application. IEEE Mirowave Magazine 5(4):50–65

Demirci MU, Nguyen CTC (2005) A low impedance VHF micromechanical filter using coupled-array composite resonators. In: Technical Digest of the 13th International Conference on Solid-State Sensors & Actuators (TRANSDUCERS), Seoul, Korea, Jun 2005, pp. 2131–2134

Demirci MU, Nguyen CTC (2006) Mechanically corner-coupled square microresonator array for reduced series motional resistance. J Microelectromech Syst 15(6):1419–1436

Esashi M (2008) Wafer level packaging of MEMS. J Micromech Microeng 18(7):073001

Fedder GK (2005) CMOS-MEMS resonant mixer-filters. In: Technical Digest of 2005 IEEE International Electron Devices Meeting, Washington, DC, Dec 2005, pp. 274–277

Fedder GK, Howe RT, Liu TJK, Quevy EP (2008) Technologies for cofabricating MEMS and electronics. Proc IEEE 96(2):306–322

Greywall DS, Busch PA (2002) Coupled micromechanical drumhead resonators with practical applications as electromechanical bandpass filters. J Micromech Microeng 12:925–938

Hao Z, Ayazi F (2007) Support loss in the radial bulk-mode vibrations of center-supported micromechanical disk resonators. Sens Actuators A 134:582–593

Hao Z, Pourkamali S, Ayazi F (2004) VHF single-crystal silicon elliptic bulk-mode capacitive disk resonators–part I: design and modeling. J Microelectromech Syst 13(6):1043–1053

Highbeam Research (2006) SiTime introduces the world's smallest and thinnest high-Q MHz resonator. http://www.highbeam.com. Accessed 20 June 2011

Ho GK, Sundaresan K, Pourkamali S, Ayazi F (2006) Temperature compensated IBAR reference oscillators. In: Proceedings of the 19th IEEE International Conference on Micro Electro Mechanical Systems, Istanbul, Turkey, Jan 2006, pp. 910–913

Hopcroft MA, Agarwal M, Park KK, Kim B, Jha CM, Candler RN, Yama G, Murmann B, Kenny TW (2006) Temperature compensation of a MEMS resonator using quality factor as a thermometer. In: Proceedings of the 19th IEEE International Conference on Micro Electro Mechanical Systems, Istanbul, Turkey, Jan 2006, pp. 222–225

Howe RT, Muller RS (1983) Polycrystalline silicon micromechanical beams. J Electrochem Soc 130:1420–1423

Hsu WT, Nguyen CTC (2002) Stiffness-compensated temperature insensitive micromechanical resonators. In: Proceedings of the 15th IEEE International Conference on Micro Electro Mechanical Systems, Las Vegas, Nevada, Jan 2002, pp. 731–734

Hsu WT, Clark JR, Nguyen CTC (2000) Mechanically temperature-compensated flexural-mode micromechanical resonators. In: Technical Digest of IEEE International Electron Devices Meeting, San Francisco, CA, Dec 2000, pp. 399–402

Huang WL (2008) Fully monolithic CMOS nickel micromechanical resonator oscillator for wireless communications. Ph.D. thesis, Electrical Engineering, University of Michigan

Huang WL, Ren Z, Nguyen CTC (2006) Nickel vibrating micromechanical disk resonator with solid dielectric capacitive-transducer gap. In: Proceedings of 2006 IEEE International Frequency Control Symposium and Exposition, Miami, Florida, Jun 2006, pp. 839–847

Huikai X, Erdmann L, Xu Z, Gabriel KJ, Fedder GK (2002) Post-CMOS processing for high-aspect ratio integrated silicon microstructures. J Microelectromech Syst 11(2):93–101

Jia G, Madou MJ (2005) Chapter 3: MEMS fabrication. In: Gad-el-Hak M (ed.) MEMS: Design and fabrication. CRC Press, Boca Raton, FL, pp. 3-1–3-214

Johnson RA (1983) Mechanical filters in electronics, Wiley Series on Filters. John Willey & Sons, New York

Kaajakari V, Mattila T, Oja A, Kiihamaki J, Seppa H (2004) Square-extensional mode single-crystal silicon micromechanical resonator for low-phase-noise oscillator applications. IEEE Electron Dev Lett 25:173–175

Kim HC, Chun K (2007) RF MEMS technology. IEEJ Trans 2:249–261

Kim B, Candler RN, Hopcroft M, Agarwal M, Park WT, Kenny TW (2007a) Frequency stability of wafer-scale film encapsulated silicon based MEMS resonators. Sens Actuators A 136:125–131

Kim B, Melamud R, Hopcroft MA, Chandorkar SA, Bahl G, Messana M, Candler RN, Yama G, Kenny T (2007b) Si-SiO2 composite MEMS resonators in CMOS compatible wafer-scale thin-film encapsulation. In: Proceedings of the IEEE International Frequency Control Symposium, Geneva, May–Jun 2007, pp. 1214–1219

Lam CS (2008) A review of the recent development of MEMS and crystal oscillators and their impacts on the frequency control products industry. In: Proceedings of the 2008 IEEE International Ultrasonics Symposium, Beijing, China, Nov 2008, pp. 694–704

Lee JEY, Seshia AA (2009) 5.4-MHz single-crystal silicon wine glass mode disk resonator with quality factor of 2 million. Sens Actuators A 156:28–35

Lee S, Demirci MU, Nguyen CTC (2001) A 10-MHz Micromechanical resonator Pierce reference oscillator for communications. In: Proceedings of the 11th International Conference on Solid State Sensors and Actuators, Munich, Germany, Jun 2001, pp. 1094–1097

Lee JEY, Bahreyni B, Zhu Y, Seshia AA (2008) A single-crystal-silicon bulk-acoustic-mode microresonator oscillator. IEEE Electron Dev Lett 29(7):701–703
Leeson DB (1966) A simple model of feedback oscillator noise spectrum. Proc IEEE 54:329–330
Li SS, Lin YW, Ren Z, Nguyen CTC (2006) Disk-array design for suppression of unwanted modes in micromechanical composite-array filters. In: Proceedings of the 19th IEEE International Conference on Micro Electro Mechanical Systems, Istanbul, Turkey, Jan 2006, pp. 866–869
Li SS, Lin YW, Xie Y, Ren Z, Nguyen CTC (2004) Micromechanical "hollow-disk" ring resonators. In: Proceedings of the 17th IEEE International Conference on Micro Electro Mechanical Systems, Maastricht, The Netherlands, Sept 2004, pp. 821–824
Lin YW, Li SS, Ren Z, Nguyen CTC (2005) Low phase noise array-composite micromechanical wine-glass disk oscillator. In: Technical Digest of 2005 IEEE International Electron Devices Meeting, Washington, DC, Dec 2005, pp. 281–284
Lin YW, Lee S, Li SS, Xie Y, Ren Z, Nguyen CTC (2004a) 60-MHz wine glass micromechanical disk reference oscillator. In: Digest of Technical Papers of 2004 IEEE International Solid-State Circuits Conference, San Francisco, CA, Feb 2004, pp. 322–323
Lin YW, Lee S, Li SS, Xie Y, Ren Z, Nguyen CTC (2004b) Series-resonant VHF micromechanical resonator reference oscillators. IEEE J Solid-State Circ 39(12):2477–2491
Lucyszyn S (2004) Review of radio frequency microelectromechanical systems technology. IEE Proc Sci Meas Technol 151(2):93–103
Mattila T, Jaakkola O, Kiihamaki J, Karttunen J, Lamminmaki T, Rantakari P, Oja A, Seppa H, Kattelus H, Tittonen I (2002) 14 MHz micromechanical oscillator. Sens Actuators A 97–98:497–502
Melamud R, Hopcroft M, Jha C, Kim B, Chandorkar S, Candler R, Kenny TW (2005) Effects of stress on the temperature coefficient of frequency in double clamped resonators. In: Digest of Technical Papers of 13th International Conference on Solid State Sensors, Actuators and Microsystems (TRANSDUCERS), Seoul, Korea, Jun 2005, pp. 392–395
Merono JD (2007) Integration of CMOS-MEMS resonators for radiofrequency applications in the VHF and UHF bands. PhD thesis, Departament d'Enginyeria Electronica, UAB
Mestrom RMC, Fey RHB, van Beek JTM, Phan KL, Nijmeijer H (2008) Modelling the dynamics of a MEMS resonator: Simulations and experiments. Sens Actuators A 142(1):306–315
Nabki F, Cicek PV, Dusatko TA, El-Gamal MN (2011) Low-Stress CMOS-Compatible Silicon Carbide Surface-Micromachining Technology—Part II: Beam Resonators for MEMS Above IC. J Microelectromech Syst 20(3):730–744
Naito Y, Helin P, Nakamura K, De Coster J, Guo B, Haspeslagh L, Onishi K, Tilmans HAC (2010) High-Q torsional mode Si triangular beam resonators encapsulated using SiGe thin film. In: Technical Digest of 2010 IEEE International Electron Devices Meeting, San Francisco, CA, Dec 2010, pp. 7.1.1–7.1.4
Nathanson HC, Newell WE, Wickstrom RA, Davis JR (1967) The resonant gate transistor. IEEE Trans Electr Dev 14(3):117–133
Nguyen CTC (1995) Micromechanical resonators for oscillators and filters. In: Proceedings of the IEEE International Ultrasonics Symposium, Seattle, WA, Nov 1995, pp. 489–499
Nguyen CTC (2007) MEMS technology for timing and frequency control. IEEE Trans Ultrason Ferroelectr Freq Control 54:251–270
Nguyen CTC (2009) Mechanical radio. IEEE Spectrum 46(12):30–35
Nguyen CTC, Howe RT (1993) CMOS micromechanical resonator oscillator. In: Technical Digest of IEEE International Electron Devices Meeting, Washington, DC, Dec 1993, pp. 199–202

Partridge A, Lutz M, Kim B, Hopcroft M, Candler RN, Kenny TW, Petersen K, Esashi M (2005) MEMS resonators: getting the packaging right. In: Proceedings of the 9th SEMI Microsystem/MEMS Seminar (SEMICON), Makuhari, Japan, Dec 2005, pp. 55–58

Pomarico A, Morea A, Flora P, Roselli G, Lasalandra E (2010) Vertical MEMS resonators for real-time clock applications. J Sensors 2010:362439

Pourkamali S (2004) Electrically coupled MEMS bandpass filters. M.S. thesis, School of Electrical and Computer Engineering, Georgia Institute of Technology

Pourkamali S, Ayazi F (2005) Electrically coupled MEMS bandpass filters Part I: With coupling element. Sens Actuators A 122:307–316

Pourkamali S, Hao Z, Ayazi F (2004) VHF single crystal silicon capacitive elliptic bulk-mode disk resonators–part II: implementation and characterization. J Microelectromech Syst 13(6):1054–1062

Pourkamali S, Hashimura A, Abdolvand R, Ho GK, Erbil A, Ayazi F (2003) High-Q single crystal silicon HARPSS capacitive beam resonators with self-aligned sub-100-nm transduction gaps. J Microelectromech Syst 12(4):487–496

Quevy EP, San Paulo A, Basol E, Howe RT, King TJ, Bokor J (2006) Back-end-of-line poly-SiGe disk resonators. In: Proceedings of the 19th IEEE International Conference on Micro Electro Mechanical Systems, Istanbul, Turkey, Jan 2006, pp. 234–237

Roessig T, Howe RT, Pisano AP (1997) Nonlinear mixing in surface-micromachined tuning fork oscillators. In: Proceedings of the 1997 IEEE Frequency Control Symposium, Orlando, FL, May 1997, pp. 778–782

Roessig TA, Howe RT, Pisano AP, Smith JH (1998) Surface-micromachined 1 MHz oscillator with low-noise Pierce configuration. In: Technical Digest of Solid State Sensors and Actuators Workshop, Cleveland, OH, Jun 1998, pp. 328–332

Schmidt MA, Howe RT (2008) Silicon resonant microsensors. In: Proceedings of Ceramic Engineering and Science, chapter 3, John Wiley & Sons, Inc., New York, pp. 1019–1034

Stephanou PJ, Pisano AP (2006) GHz contour extensional mode aluminum nitride MEMS resonators. In: Proceedings of the 2006 IEEE Ultrasonics Symposium, Berkeley, CA, Oct 2006, pp. 2401–2404

Sundaresan K, Ho GK, Pourkamali S, Ayazi F (2005) A two-chip, 4-MHz, microelectromechanical reference oscillator. In: Proceeding of the 2005 IEEE International Symposium on Circuits and Systems, Kobe, Japan, May 2005, vol. 6, pp. 5461–5464

Sworowski M, Neuilly F, Legrand B, Summanwar A, Lallemand F, Philippe P, Buchaillot L (2009) High-Q and low-Rm 24-MHz radial-contour mode disk resonators fabricated with silicon passive integration technology. In: Proceedings of the 15th IEEE International Conference on Solid-State Sensors, Actuators and Microsystems (TRANSDUCERS), Denver, CO, Jun 2009, pp. 2114–2117

Sze SM (2003) VLSI technology, 2nd edn. Tata McGraw Hill, New Delhi

Tabatabaei S, Partridge A (2010) Silicon MEMS oscillators for high-speed digital systems. IEEE Micro 30(2):80–89

Takeuchi H, Quevy E, Bhave SA, King TJ, Howe RT (2004) Ge-blade damascene process for post-CMOS integration of nano-mechanical resonators. IEEE Electron Dev Lett 25:529–531

Tang WC, Nguyen TCH, Howe RT (1989) Laterally driven polysilicon resonant microstructures. In: Proceedings of IEEE Micro Electro Mechanical Systems, Salt Lake City, UT, Feb 1989, pp. 53–59

Taylor JT, Huang Q (1997) CRC Handbook of Electrical Filters. CRC Press, Boca Raton, FL

Tilmans HAC (1996) Equivalent circuit representation of electromechanical transducers: I, Lumped-parameter systems. J Micromech Microeng 6:157–176

Tilmans HAC, Raedt WD, Beyne E (2003) MEMS for wireless communications: 'from RF-MEMS components to RF-MEMS-SiP'. J Micromech Microeng 13: S139–S163

Tummala RR, Swaminathan M (2008) Introduction to system-on-package (SOP): miniaturization of the entire system. McGraw-Hill Professional, New York

Vittoz EA, Degrauwe MGR, Bitz S (1988) High-performance crystal oscillator circuits: theory and application. IEEE J Solid-State Circuits 23:774–783

Wang K, Nguyen CTC (1999) High-order medium frequency micromechanical electronic filters. J Microelectromech Syst 8(4):534–557

Wang K, Wong AC, Nguyen CTC (2000) VHF free-free beam high-Q micromechanical resonators. J Microelectromech Syst 9(3):347–360

Wang J, Ren Z, Nguyen CTC (2004a) 1.156-GHz self-aligned vibrating micromechanical disk resonator. IEEE Trans Ultrason Ferroelectr Freq Control 51:1607–1628

Wang J, Butler JE, Feygelson T, Nguyen CTC (2004b) 1.51-GHz Nanocrystalline diamond micromechanical disk resonator with material-mismatched isolating support. In: Proceedings of 17th IEEE Int. Conf. on Micro Electro Mechanical Systems, Maastricht, The Netherlands, Sept 2004, pp. 641–644

Wong AC, Nguyen CTC (2004) Micromechanical mixer-filters ("Mixlers"). J Microelectromech Syst 13:100–112

Wong TSA, Palaniapan M (2009) Micromechanical oscillator circuits: theory and analysis. Analog Integr Circ Sig Process 59:21–30

Wong AC, Clark JR, Nguyen CTC (1999) Anneal-activated, tunable, 65MHz micromechanical filters. In: Digest of Technical Papers of 10th International Conference on Solid-State Sensors and Actuators, Sendai, Japan, Jun 1999, pp. 1390–1393

Yole (2010) Emerging MEMS: Technologies & Markets, 2010 Report. Yole Développement, France

Zuo C, Van der Spiegel J, Piazza G (2010) 1.05-GHz CMOS oscillator based on lateral-field-excited piezoelectric AlN contour-mode MEMS resonators. IEEE Trans Ultrason Ferroelectr Freq Control 57:82–87